\documentclass{pasa}%
\def\msun{{\Msun}} 
\def\Msun{\rm M_\odot}
\def\lya{{\rm Ly}\alpha}
\def\HEII{\hbox{He~$\scriptstyle\rm II$~}} 
\def\lum{{\rm erg}\,{\rm s}^{-1}}

\newcommand{\newqso}
  {ULAS~J1120+0641}

\newcommand\kms{km~s$^{-1}$}
\title[Physical properties of the first quasars]{Physical properties of the first quasars}
\author[Gallerani et al.]{S. Gallerani$^1$\thanks{simona.gallerani@sns.it}, X. Fan$^2$, R. Maiolino$^{3,4}$, F. Pacucci$^{1,5}$\\
\affil{$^1$ Scuola Normale Superiore, Piazza dei Cavalieri 7, 56126, Pisa, Italy} 
\affil{$^2$ Steward Observatory, University of Arizona, 933 N Cherry Ave., Tucson, AZ, 85721, USA}
\affil{$^3$ Kavli Institute for Cosmology, University of Cambridge, Madingley Road, Cambridge CB3 0HA, United Kingdom}
\affil{$^4$ Cavendish Laboratory, University of Cambridge, 19 J. J. Thomson Ave., Cambridge CB3 0HE, United Kingdom} 
\affil{$^5$ Department of Physics, Yale University, P.O. Box 208121, New Haven, CT 06520, USA}}
\jid{PASA}
\doi{10.1017/pas.\the\year.xxx}
\jyear{\the\year}

\usepackage[authoryear]{natbib}
\usepackage{enumitem}   
\bibpunct{(}{)}{;}{a}{}{,}
\usepackage{aas_macros}
\usepackage{hyperref} 
\hypersetup{colorlinks,citecolor=blue,linkcolor=blue,urlcolor=blue}

\begin{document}%
\begin{abstract}
Since the beginning of the new millennium, more than 100 $z\sim 6$ quasars have been discovered through several surveys and followed-up with multi-wavelength observations. These data provided a large amounts of information on the growth of supermassive black holes at the early epochs, the properties of quasar host galaxies and the joint formation and evolution of these massive systems. We review the properties of the highest-$z$ quasars known so far, especially focusing on some of the most recent results obtained in (sub-)millimeter bands. 
We discuss key observational challenges and open issues in theoretical models and highlight possible new strategies to improve our understanding of the galaxy-black hole formation and evolution in the early Universe.
\end{abstract}
\maketitle%
\section{INTRODUCTION }
\label{sec:intro}
\begin{figure*}
\centering
\includegraphics[angle=0,width=0.9\textwidth]{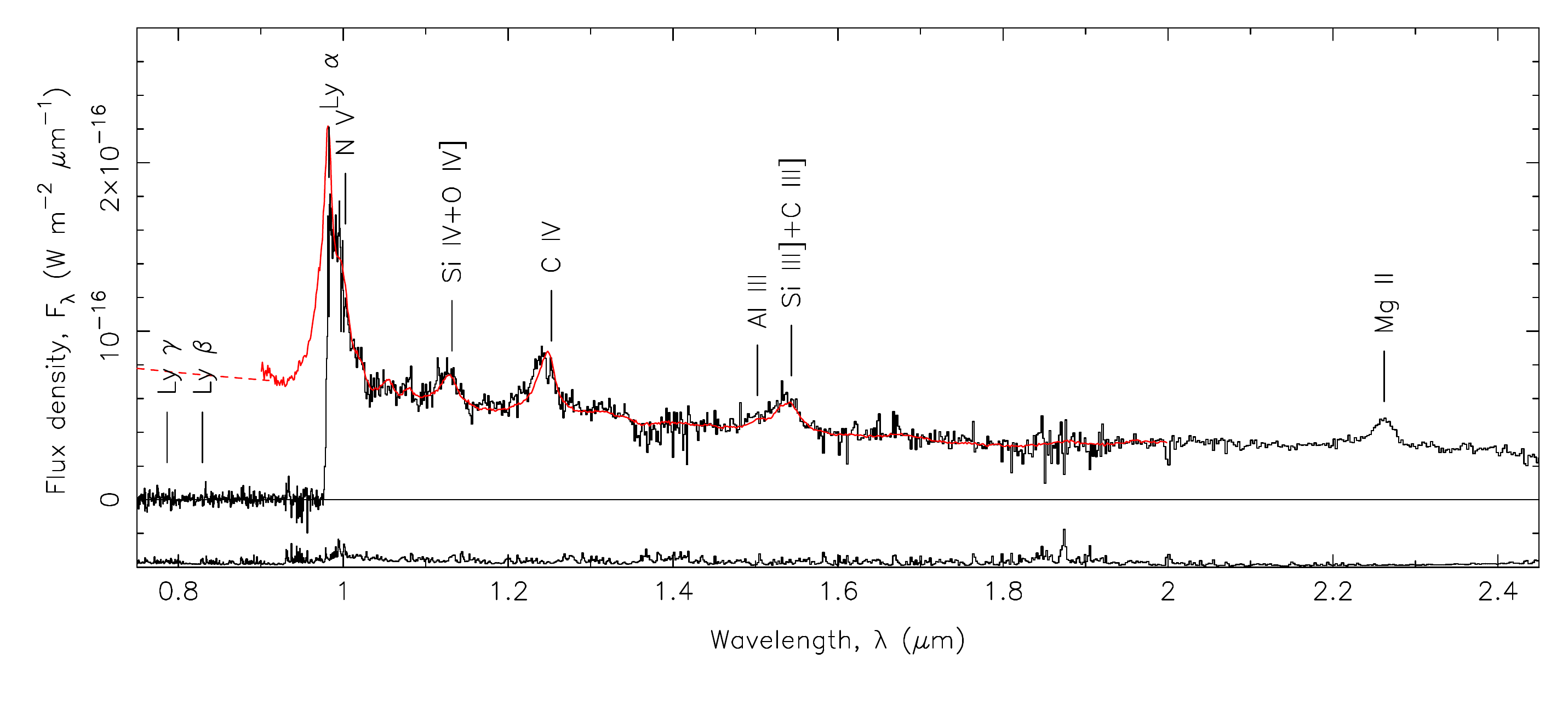}
\caption{Spectrum of \newqso\ (black line) compared to a composite spectrum derived from lower-redshift quasars (red line, Telfer et al. 2002). Adapted from Fig. 1 of Mortlock et al. (2011) by permission of the authors and the Nature Publishing Group; courtesy of Daniel Mortlock.}
\label{figure:spectrum}
\end{figure*}
\begin{figure}
\centering
\includegraphics[angle=0,width=0.52\textwidth]{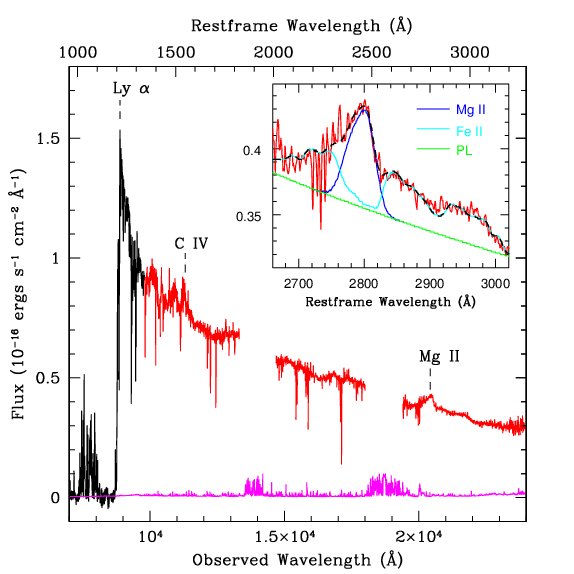}
\caption{Combined optical/near-infrared spectrum of J0100+2802. Based on the Mg II Full Width at Half Maximum (FWHM=$5130\pm150~{\rm km~s}^{-1}$), and the continuum luminosity at the rest-frame wavelength of 3000~\AA~($3.15\pm0.47 \times 10^{47} {\rm ergs~s}^{-1}$), Wu et al. (2015) estimate a black hole mass $(1.24\pm 0.19) \times 10^{10} M_{\odot}$ for this source. Reproduced from Figure 3 of Wu et al. (2015) by permission of the authors and the Nature Publishing Group.}
\label{figure:mostlum}
\end{figure}
Multi-wavelength data of $z\sim 6$ quasars have shown to be fundamental to understand the properties of the high redshift Universe both on cosmological ($\geq$ tens of Mpc) and galactic scales ($\leq$ tens of kpc). In the last two decades, more than 100 $z\sim 6$ quasars have been discovered through various different surveys. Initially, most of them were found at $5.7 \leq z \leq 6.4$ with the SDSS and the CFHQS (e.g. Jiang et al. 2009, Willott et al. 2010 and references therein). Nowadays, new surveys with enhanced sensitivities in the red and near-IR as Pan-STARRS1 (Ba$\rm \tilde{n}$ados et al. 2016), VST/ATLAS (Carnall et al. 2015), DES (Reed et al. 2015, 2017), Subaru/HSC (Matsuoka et al. 2016), UKIDSS (Venemans et al. 2007) and VIKING (Venemans et al. 2013, 2015a) are discovering further quasars at similar and higher redshift. Currently, the highest redshift quasar known is ULASJ1120+0641, discovered in UKIDSS at z = 7.09 (Mortlock et al. 2011, Fig. \ref{figure:spectrum}).

On large scales, observations of $z\sim 6$ quasars have been extensively used as powerful probes of the cosmic reionization process.
Optical/near-infrared absorption spectra of $z\sim 6$ quasars in the spectral region blue-ward the Ly$\alpha$ emission line are characterized by the presence of ionized near zones (2-12 physical Mpc, e.g. Fan et al. 2006; Venemans et al. 2015b; Reed et al. 2017) interrupted by deep absorption gaps that extend up to hundreds of Mpc (Gallerani et al. 2006; McGreer et al. 2015; Becker et al. 2015). The sizes of quasar near zones are generally used to constrain the intergalactic medium (IGM) neutral hydrogen fraction (e.g. Mesinger \& Haiman 2004; Maselli et al. 2007; Keating et al. 2015). In particular, evidence for a Ly$\alpha$ damping wing has been found  in ULASJ1120+0641 suggesting a neutral hydrogen fraction $x_{\rm HI}>0.1$ (Mortlock et al. 2011). More recently, Greig et al. (2016a) have used the same observations to provide more stringent constraints on the neutral hydrogen fraction ($x_{\rm HI}=0.4^{+0.21}_{-0.19}$) thanks to the Ly$\alpha$ emission line reconstruction method presented by Greig et al. (2016b). These studies on the IGM ionization state are fundamental to constrain models of the cosmic reionization process (e.g. Choudhury \& Ferrara 2005; Choudhury et al. 2008; Mitra et al. 2012; Mitra et al. 2016) to finally understand this important phase transition in the evolution of the Universe. Ionized regions around high-$z$ quasars also provide measurements of the average IGM temperature at the mean density ($T_0=10^{4.21\pm 0.03}\rm K$, Bolton et al. 2012), the quasar lifetimes ($t_Q\sim 10^6-10^8 \rm yr$, e.g Gallerani et al. 2008) and the environment in which high-$z$ quasars form and evolve (e.g. Kim et al. 2009; Maselli et al. 2009). 

On small scales, $z\sim 6$ quasars are ideal laboratories for studying the properties of the host galaxy and its coeval formation with the super-massive black holes (SMBHs) they contain. In the spectral region red-ward the Ly$\alpha$ emission line, the quasar continuum retains precious information on the dust properties of the host galaxy. It has been shown that extinction properties of high-$z$ quasars are well described in terms of SN TypeII dust (e.g. Maiolino et al. 2004; Gallerani et al. 2010), though this argument is still subject of hot debate. Moreover, while metal absorption lines can be used to study the global IGM metal enrichment (e.g. Pallottini et al. 2014; D'Odorico et al. 2013; Oppenheimer et al. 2012; Ryan-Weber et al. 2009), metal emission lines provide information on the metallicity of the quasar host galaxy (e.g. Fan et al. 2004; Juarez et al. 2009). The inferred metallicity of the gas in the quasar broad line region is super-solar, similarly to what is observed at lower redshifts. Moreover, MgII emission line observations in the near-IR provide measurements of the mass of black holes powering $z\sim 6$ quasars (e.g. Willott et al. 2003). These studies show that high-$z$ quasar host galaxies contain SMBHs rapidly grown up to a mass $M_{\bullet}\geq 10^{9}M_{\odot}$. The current most massive black hole ($M_{\bullet}=1.24 \pm 0.19 \times 10^{10}M_{\odot}$) is powering a quasar at $z=6.3$ (Wu et al. 2015, Fig. \ref{figure:mostlum}). The presence of SMBHs formed in such a short time strongly challenges the standard black hole growth theory, as extensively discussed in Sec. \ref{SMBHs}. The detection of SMBH progenitors would be extremely important both to understand the formation of these massive systems (e.g. Pezzulli et al. 2016, 2017) and to clarify the contribution of faint quasars to the cosmic reionization process (e.g. Volonteri \& Gnedin 2009; Giallongo et al. 2015; Madau \& Haardt 2015; Manti et al. 2017; Kulkarni et al. 2017).

More recently, radio, millimeter and sub-millimeter observations (VLA, ALMA and NOEMA) have provided a new window on $\leq$kpc scales to study the cold gas and dust physical condition in the interstellar medium (ISM) of $z\sim 6$ quasar host galaxies (see Carilli \& Walter 2013 and references therein). These observations have been useful to characterize the properties of $z\sim 6$ quasar host galaxies (e.g. in terms of star formation rate (SFR), dust masses, dynamical mass, molecular hydrogen mass, feedback), to understand the co-evolution of galaxies and their SMBHs at high-$z$ (e.g. Valiante et al. (2017) for a recent theoretical review on this subject), and may provide new tools for detecting SMBH progenitors with important implications for the formation of these massive systems and for the contribution of faint quasars to cosmic reionization. 

In this review, we first discuss the latest results obtained from studies on the SMBH formation; then, we focus on the most recent millimeter and sub-millimeter observations of far infrared (FIR) emission lines arising from the ISM of $z\sim 6$ quasars; finally, we discuss possible observational strategies that could help us progressing in the comprehension of the first quasar physical properties.
\section{The formation and evolution of SMBHs}\label{SMBHs}
Black hole mass estimates rely on the measurement of broad emission lines (BELs) widths, and are based on the assumption that the dynamics of the broad line regions (BLRs) are dominated by the gravity of the central BH (see Peterson et al. 2014). The black hole mass can thus be expressed as:
\begin{equation}\label{virial_product}
 M_{\rm BH}\propto G^{-1}~R_{\rm BLR}v_{\rm BLR}^2,  
\end{equation}
where G is the gravitational constant, $R_{\rm BLR}$ is the BLR radius (or, stated differently, the distance of the line-emitting gas from the central ionizing source) and $v_{\rm BLR}^2$ is the BLR velocity dispersion (related to the FWHM of the BELs). Reverberation mapping of H$\beta$ emission in local active galactic nuclei (AGNs) has shown that a relationship exists between the AGN optical/UV luminosity and $R_{\rm BLR}$(e.g. Bentz et al. 2013). This result, combined with eq. \ref{virial_product}, implies that the measurement of the H$\beta$ FWHM and the optical/UV luminosity are sufficient to determine $M_{\rm BH}$. However, the H$\beta$ emission line is redshifted out of the optical window at $z\sim 0.9$ and above; in this cases $M_{\rm BH}$ estimates are then based on the CIV and MgII emission lines (Vestergaard \& Peterson 2006). These methods typically provide a scatter of 0.15, 0.21, and 0.27 dex for the MgII, H$\beta$, and CIV lines, respectively (Steinhardt \& Elvis 2010). We finally note that recent works by Brotherton et al. (2015) and Mejia-Restrepo et al. (2016) have found that virial $M_{\rm BH}$ estimates obtained from the CIV line are not reliable, possibly indicating that the gas emitting this line is not virialized.

The maximum black hole masses found at high-redshifts are comparable to the most massive objects found in the local Universe and to the theoretical maximum black hole mass achievable by luminous accretion ($\sim 10^{10} \mathrm{M_{\odot}}$, see Natarajan \& Treister 2009, King 2016). The challenge is to understand how these SMBHs have formed in less than $\sim$1 Gyr, the age of the Universe at $z \sim 6$ (e.g. Volonteri et al. 2003; Volonteri 2010). 
 
In the standard growth scenario, the luminosity of an accreting black hole can be expressed through the following relation:
\begin{equation}
L=c^2\frac{\epsilon}{1-\epsilon}\dot{M}_{\rm BH},
\end{equation}
where $\epsilon\sim 0.1$ is the efficiency at which accreting gas rest mass is converted in radiation, and $\dot{M}_{\rm BH}$ is the black hole mass accretion rate (see Trakhtenbrot et al. 2017a for a recent work on radiative efficiencies and accretion rates of $z\sim 6$ quasars). Under the assumption that the black hole is shining at the Eddington luminosity ($L_E=\frac{4\pi cGm_p}{\sigma_T}$), it grows in mass exponentially:
\begin{equation}
M_{\rm BH}(t)=M_{\rm BH}(0)~\rm exp\left(\frac{1-\epsilon}{\epsilon}\frac{t}{t_{\rm Edd}}\right),
\end{equation}
where $t_{\rm Edd}=0.45$~Gyr is the Eddington time.
This formula can be used to estimate the minimum mass of the BH seed at a given time $M_{\rm BH}(0)$ such that it evolves into a SMBH at $z\sim 6$. 

Black hole seeds can either form small and grow fast (i.e. overshooting the Eddington limit), or form big and grow at a normal pace, at or below Eddington.
In the first scenario, BH seeds are formed from the collapse of primordial (PopIII) stars, metal-free objects as massive as hundreds of solar mass and hosted in dark matter mini-halos ($M_h\sim 10^{5-6}~M_{\odot}$) at $z>20$ (e.g. Tegmark et al. 1997; Abel et al. 2002; see also Ciardi \& Ferrara for a review on this topic). 
BH remnants from PopIII stars are expected to be $\leq 100 \, \mathrm{M_{\odot}}$ (e.g. Schneider et al. 2002; Zhang et al. 2008).  Although this path to form a BH seed seems to be very natural, large uncertainties exist on the final mass of PopIII stars and only the most optimistic assumptions can explain the SMBH origin from PopIII remnants. The feasibility of this scenario could be improved by allowing accretion rates not to be limited to the Eddington rate (see Volonteri \& Rees 2005, Alexander \& Natarajan 2014, Madau et al. 2014, Volonteri et al. 2015). These works suggest that short and recurring episodes of super-Eddington accretion may take place, speeding up the growth process. This may occur in heavily buried seeds, in which photon trapping reduces the efficiency of radiation pressure.
\begin{figure*}
\centering
\includegraphics[angle=0,width=0.50\textwidth]{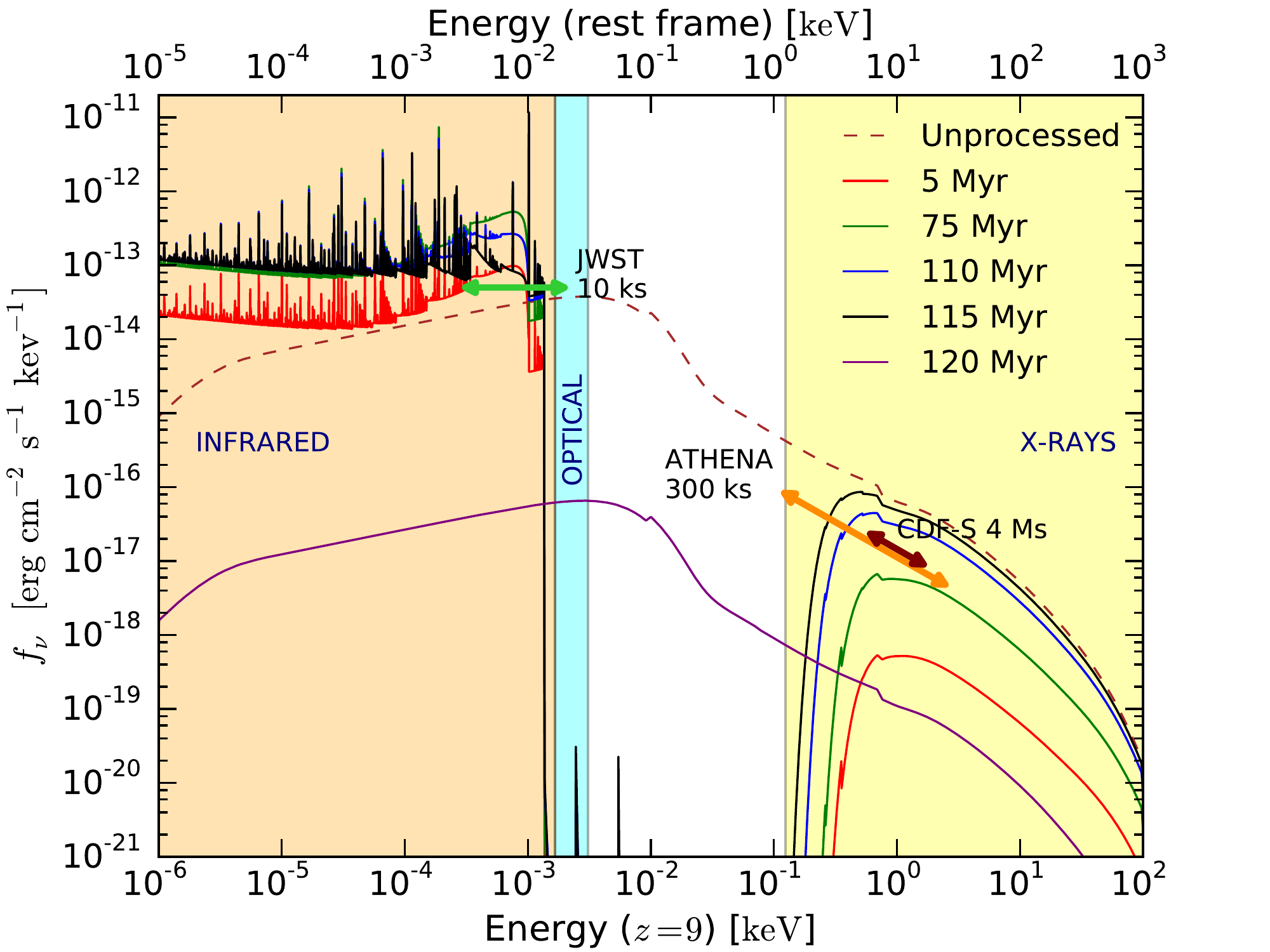}
\includegraphics[angle=0,width=0.49\textwidth]{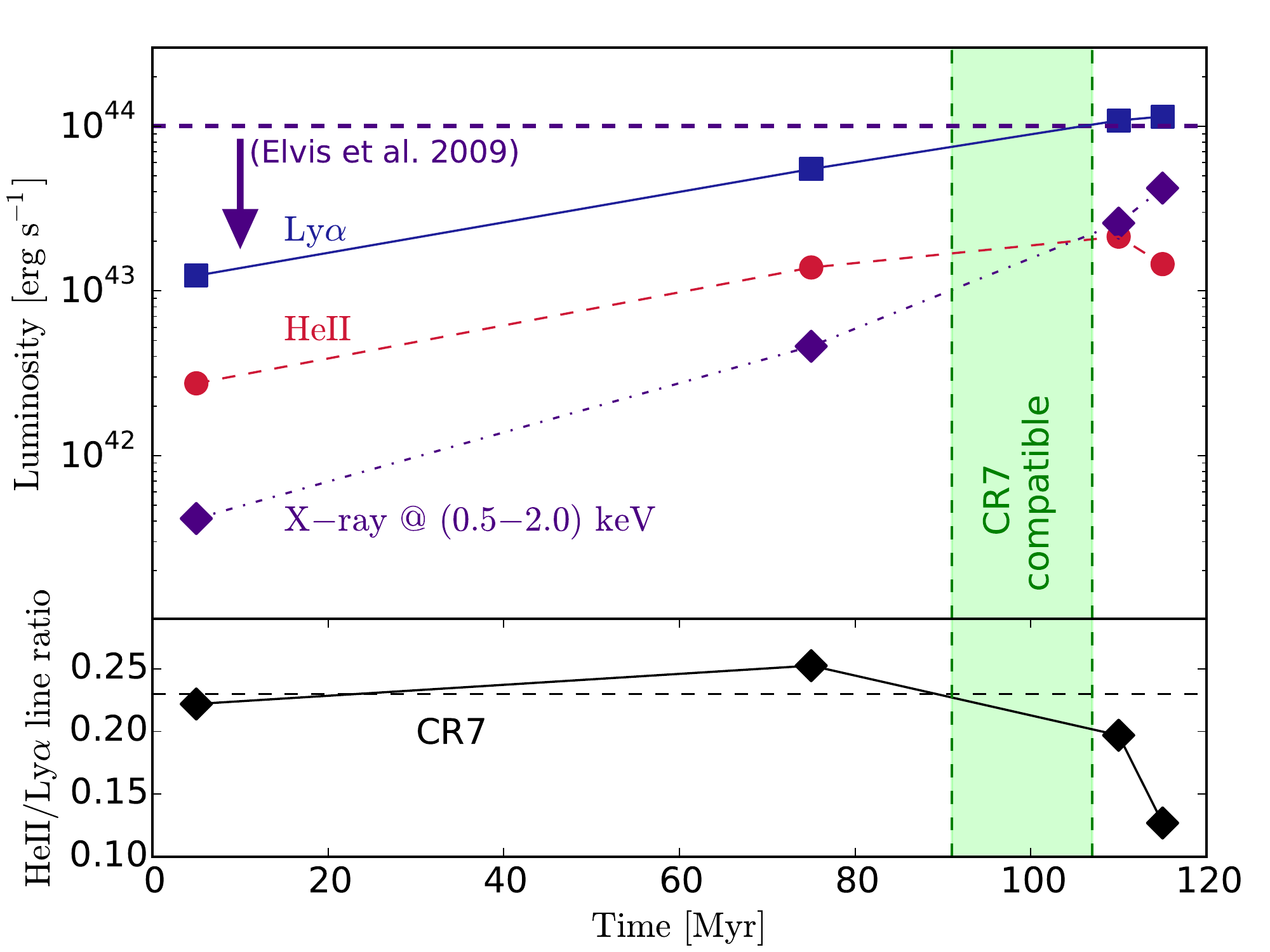}
\caption{{\bf Left panel}: Time evolution of the spectrum emerging from the host halo for a source located at $z=9$, in the standard accretion case, with the density profile investigated in Pacucci et al. (2015a), i.e. with a core number density of $10^6$ cm$^{-3}$. The infrared, optical and X-ray bands are highlighted with shaded regions, while the unprocessed spectrum is reported, at peak luminosity ($t=115 \, \mathrm{Myr}$), with a dashed line. The flux limits for future (JWST, ATHENA) and current (CDF-S) surveys are also shown. The left panel is reproduced from Pacucci et al. (2015b). {\bf Upper right panel}: Time evolution of the $\lya$ (blue solid line), \HEII (red dashed line) line and X-ray (violet dot dashed line) luminosities calculated for the accretion process onto a DCBH of initial mass $10^5 \msun$. The green shaded region indicates the period of time during which the simulation results are compatible with CR7 observations. The current upper limit for X-ray is $\lesssim 10^{44} \, \lum$, (horizontal violet line, Elvis et al. 2009). {\bf Lower right panel}:
Time evolution of the ${\rm He II}/\lya$ lines ratio. The black horizontal dashed line indicates the observed values for CR7. The right panel is reproduced from Pallottini et al. (2015).
\label{fig_dcbh}}
\end{figure*}
In the alternative scenario, black hole seeds are formed with heavy masses, from $\sim 10^3 \, \mathrm{M_{\odot}}$ up to $\sim 10^5 \, \mathrm{M_{\odot}}$. In this case, super-Eddington accretion would not be required.
For instance, as soon as the gas is polluted by metals created in the first PopIII stars, the normal PopII star formation mode can proceed. This first episode of efficient star formation can foster the formation of very compact nuclear star clusters where star collisions can lead to the formation of a very massive star, possibly leaving BH remnants with masses $\sim 10^3 \, \mathrm{M_{\odot}}$ (e.g. Devecchi \& Volonteri 2009, Davies et al. 2011). 
Even larger BH seeds may be achieved in the Direct Collapse Black Hole (DCBH) scenario, described in the following section.
\subsection{Direct Collapse Black Holes}
\begin{figure}
\vspace{-1\baselineskip}
\hspace{-0.5cm}
\begin{center}
\includegraphics[angle=0,width=0.5\textwidth]{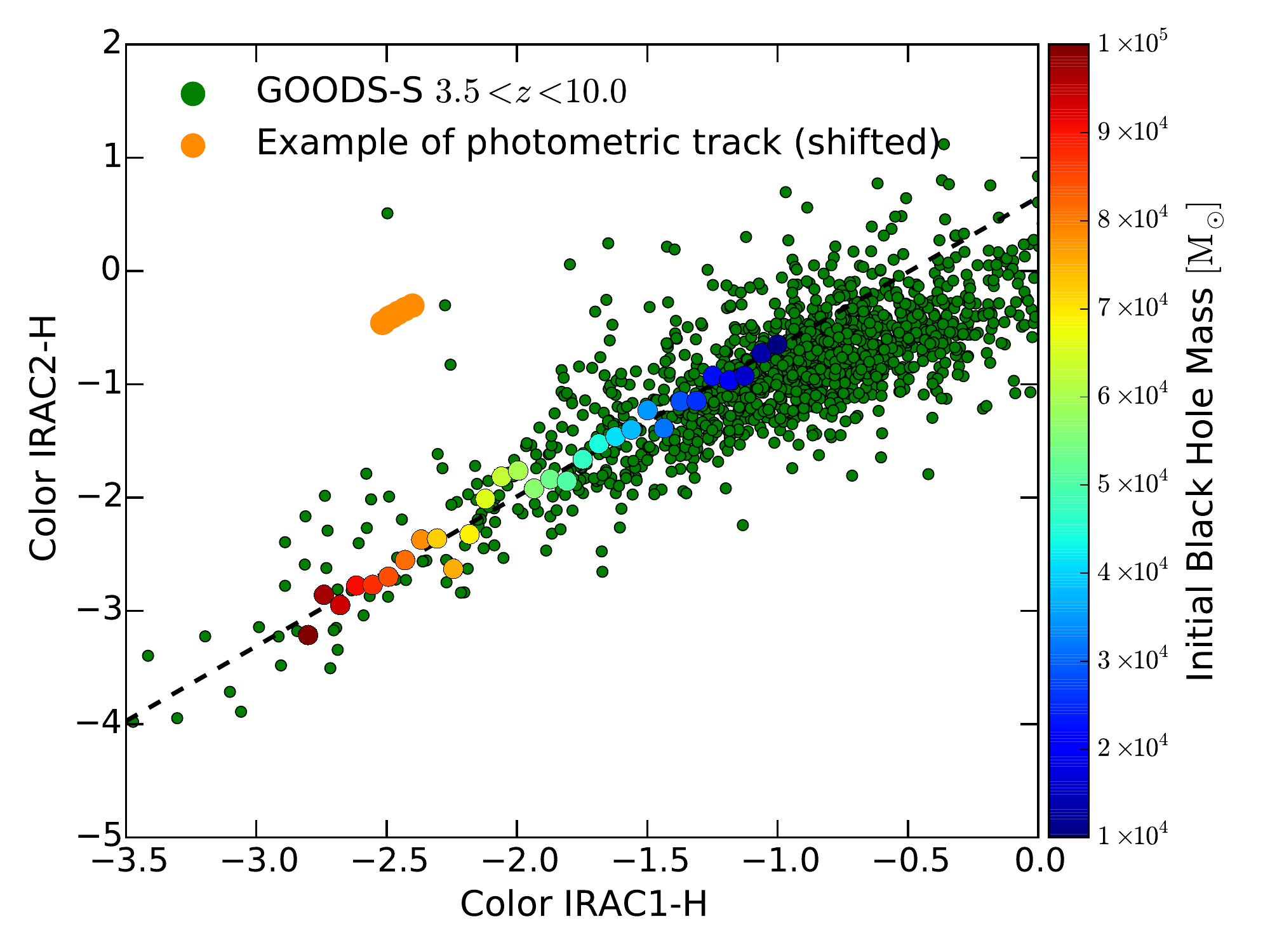}
\caption{Color-color diagram for the infrared filters H, IRAC1 and IRAC2. GOODS-S objects, brighter than the 27th magnitude in the H band ($\mathrm{H<27}$) and with $3.5 \lesssim z \lesssim 10$, are shown with green points. Numerical predictions for the colors of DCBHs are shown, at $z \sim 7$, with filled circles, whose color depends on the initial mass of the seed (see the color-bar). Larger black hole masses are associated with redder spectra (i.e. more negative colors). All colors are \textit{observed quantities}. An example of a photometric track for a DCBH of initial mass $\sim 8 \times 10^{4} \, \mathrm{\Msun}$ is shown in orange. Its position has been shifted vertically to avoid information overload. Reproduced from Pacucci et al. (2016).}
\label{fig:colors_total}
\end{center}
\end{figure}
\begin{figure*}
\includegraphics[width=8.5cm]{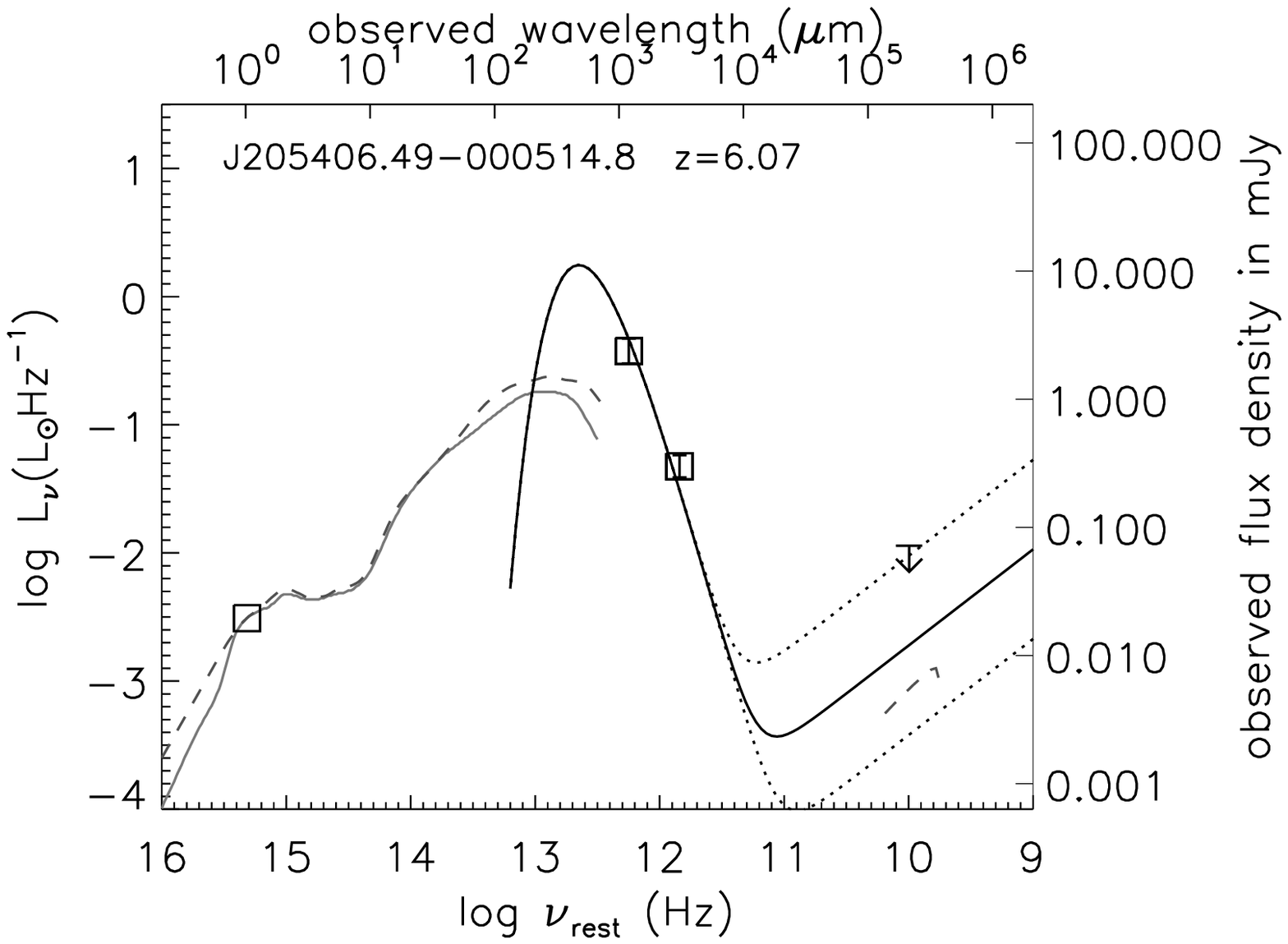}
\includegraphics[width=8.5cm]{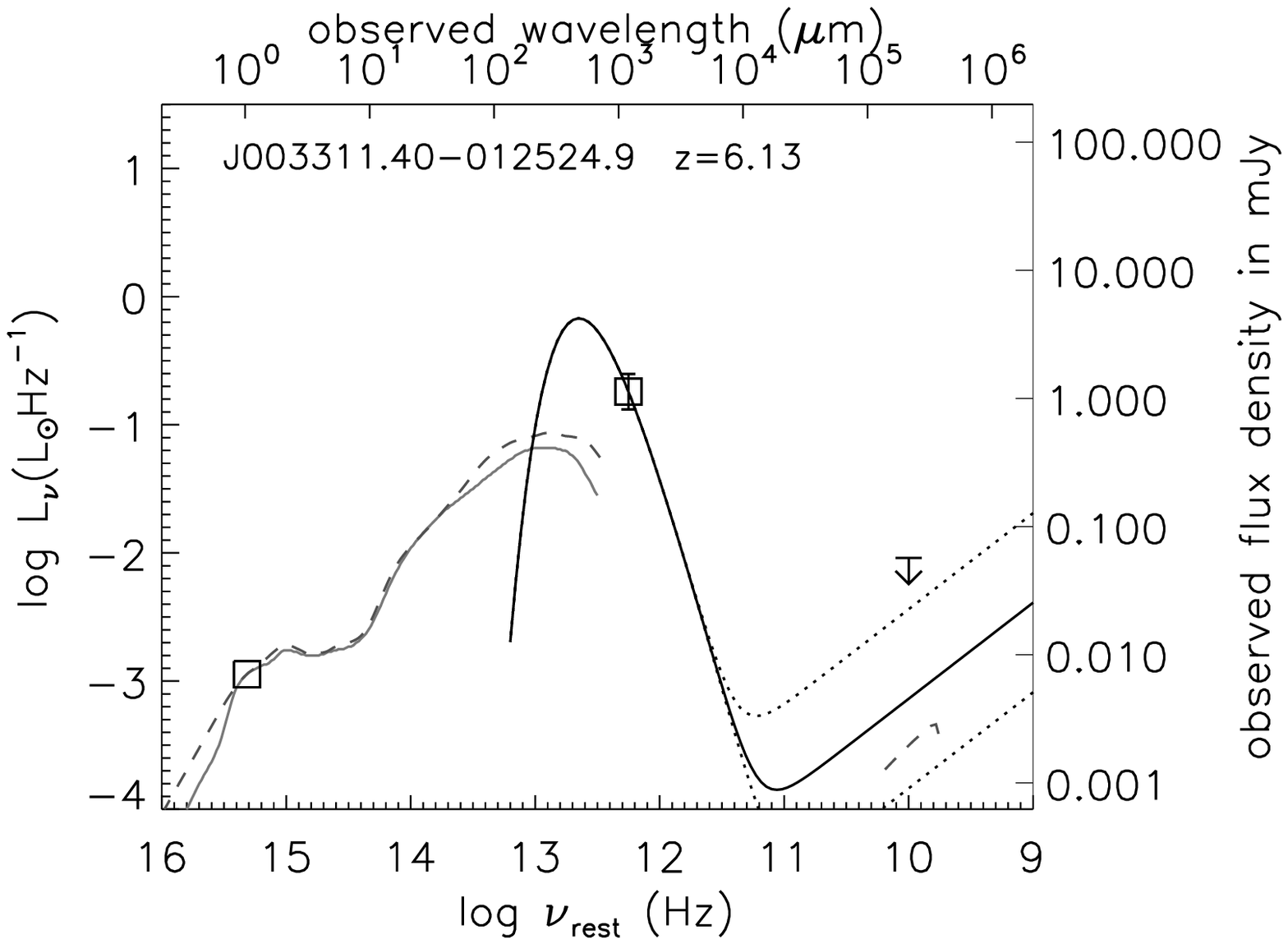}\\
\includegraphics[width=8.5cm]{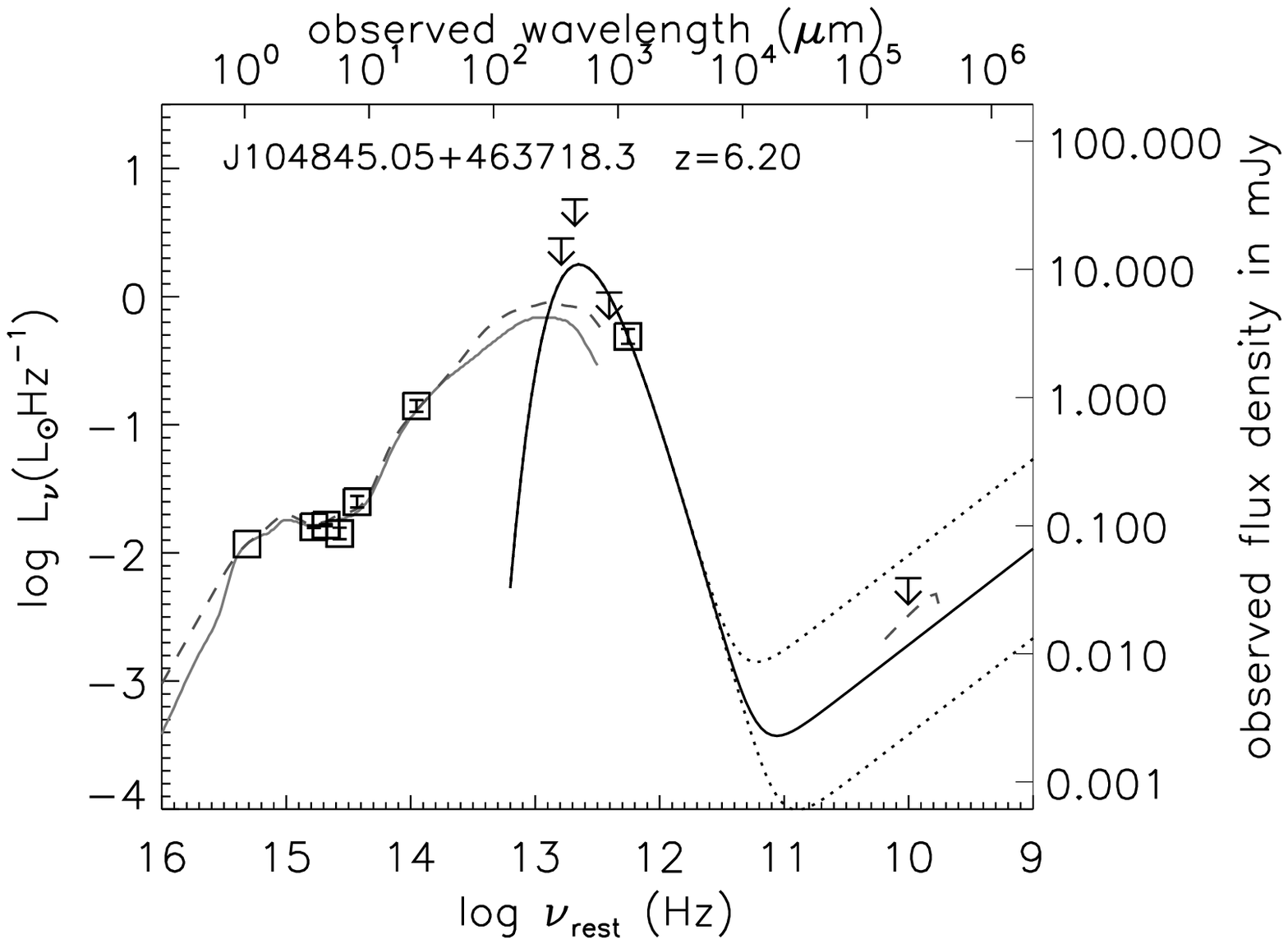}
\includegraphics[width=8.5cm]{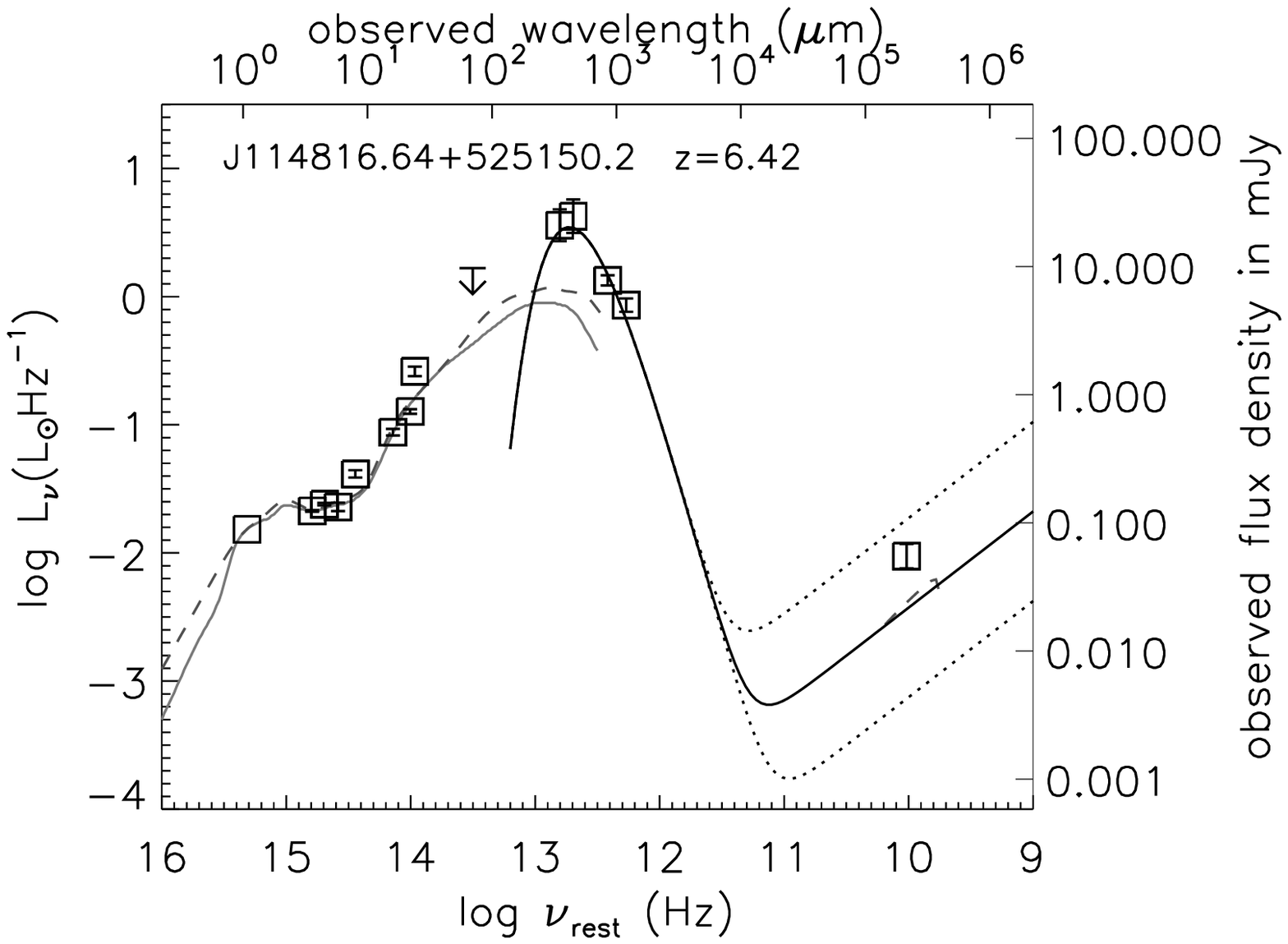}\\
\caption{Examples of UV to radio SEDs of the 250 GHz detected $z\sim 6$ quasars. The solid and dashed grey lines show local quasar templates normalized to rest frame 1450~$\rm \AA$. The thick solid line is a warm dust model normalized to the submm data and extended to the radio band with the typical radio-FIR correlation of star forming galaxies ($q=2.34$; the dotted grey lines take into account factors of 5 excesses above and below the typical $q$ value). Adapted from Fig. 3 of Wang et al. (2008) and reproduced by permission of the AAS.}
\label{fig:WangSED}
\end{figure*}
\begin{figure}
\hspace{-0.8cm}
\includegraphics[width=9.5cm]{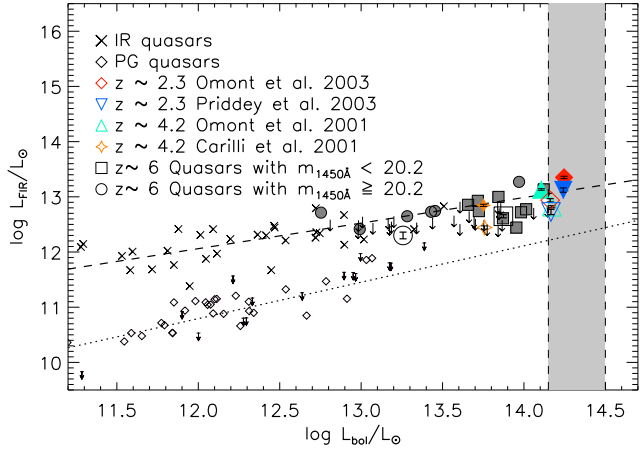}
\caption{The FIR ($8<\lambda_{\rm RF}<1000~\mu m$) and bolometric quasars luminosity correlations of quasar samples from local to $z\sim6$. The filled triangles and diamonds are the mean FIR luminosities for the
(sub)mm detections in each sample, derived by averaging flux densities at 250 GHz or 350 GHz. The open symbols
represent the mean luminosities averaged with both the millimeter detections and non-detections in each sample. The
dotted line is a fit to the local optically selected PG quasars and the dashed
line shows the fit to the submillimeter or millimeter detected sources in all high-$z$ samples and the local ULIRGs. Reproduced from Fig. 4 of Wang et al. (2011a) by permission of the AAS.}
\label{fig:WangFIR}
\end{figure}
In the early Universe, the collapse of a primordial atomic-cooling halo, exposed to a Lyman-Werner (LW, energy $h\nu =11.2-13.6 \, \mathrm{eV}$) flux intense enough to efficiently dissociate molecular hydrogen\footnote{The minimum LW flux (J$_{\rm crit}$) required to dissociate molecular hydrogen depends on several factors, but there is a general consensus that it should fall in the range $30 < J_{\rm crit}(21) < 1000$, depending on the spectrum of the sources (Sugimura et al. 2014), where J$_{\rm crit}(21)$ is the critical value of J in units of $10^{-21} \rm erg s^{-1}cm^{-2}Hz^{-1} sr^{-1}$.}, may lead to the formation of a DCBH, through a general relativistic instability (Shang et al. 2010, Johnson et al. 2012, Ferrara et al. 2014).
The rationale of these conditions is that the primordial gas needs to contract without cooling and fragmenting into stars. Then, the absence of the main coolants (metals and molecular hydrogen, dissociated by the LW radiation) may serve the purpose.
The final result of this process is the formation of a BH seed having mass of the order of $10^4-10^6 ~\mathrm{M_{\odot}}$ (see the birth mass function for DCBHs presented in Ferrara et al. 2014). 

In recent years, the scientific community has focused much attention on this scenario, likely due to the following reasons: (i) it is theoretically elegant and well tailored to the physical conditions of the early Universe, (ii) the (predicted) observational signatures of these objects are easier to check in actual observations, e.g. the absence of metal lines, and (iii) the predicted masses are sufficiently large to allow for their direct observation with current or upcoming surveys.

Much effort has been devoted recently in understanding the observational properties of these sources.
For instance, Pacucci et. al (2015b) investigated the time-evolving spectrum of an accreting DCBH of initial mass $M_{\bullet} = 10^5 \Msun$ (Fig. \ref{fig_dcbh}, left panel) by coupling a 1D radiation-hydrodynamic code (Pacucci \& Ferrara 2015) to the spectral synthesis code {\it CLOUDY} (Ferland et al. 2013). The DCBH is at the center of a halo of total gas mass $M_{g} \simeq 10^7 \msun$, distributed in a core plus an $r^{-2}$ density profile. The accretion is followed until complete depletion of the halo gas, i.e. for $\simeq 120 \, {\rm Myr}$. During this period the total absorbing column density of the gas varies from an initial value of $\sim 10^{24} {\rm cm}^{-2}$ to a final value $\ll 10^{22} {\rm cm}^{-2}$, i.e. from being mildly Compton-thick to strongly Compton-thin. Note that while $\lya$ attenuation by the interstellar medium is included, this work does not account for the likely sub-dominant IGM analogous effect. The bulk of the emission occurs in the observed infrared-submm ($1-1000 \, \mathrm{\mu m}$) and X-ray ($0.1 - 100 \, \mathrm{keV}$) bands. This work has recently been extended by Natarajan et al. (2016), with a more comprehensive study of the gas metallicity effects and seed formation mechanism on the emerging spectrum. Moreover, the first multi-wavelength spectral predictions for JWST observability of these sources were presented.

The DCBH template spectrum presented by Pacucci et al. (2015b) has been used to interpret observations of CR7 at $z\approx 6.6$ (Sobral et al. 2015). This source is the brightest Ly$\alpha$ emitter discovered to date and shows some peculiar observational features, like strong Ly$\alpha$ and HeII emission lines and absence of metal lines. Though initially this source was identified as a possible PopIII galaxy, Pallottini et al. (2015) have discarded this hypothesis (unless under somehow extreme conditions; see also Yajima \& Khochfar 2017 and Visbal et al. 2017) and suggested that this source most likely host a DCBH (see also Agarwal et al. 2016; Dijkstra et al. 2016; Smith et al. 2016). The right panel of Fig. \ref{fig_dcbh} shows the time evolution of the $\lya$, \HEII and X-ray (0.5-2 keV) luminosities for a typical DCBH of initial mass $\sim 10^5 \, \mathrm{M_{\odot}}$. Both $\lya$ and \HEII are consistent with the observed CR7 values during an evolutionary phase lasting $\simeq 17\, {\rm Myr}$ (14\% of the system lifetime).

The equivalent width of the \HEII line in the CR7 compatibility region (green shaded region in the right panel of Fig. \ref{fig_dcbh}) ranges from 75 to 85~\AA. The column density during the CR7-compatible period is $\simeq 1.7 \times 10^{24}\mathrm{cm^{-2}}$, i.e. mildly Compton-thick. The associated X-ray luminosity is $L_X\lesssim 10^{43}\lum$, fully consistent with the current upper limit for CR7 ($L_X\lesssim 10^{44}\lum$) obtained by Elvis et al. (2009). Deeper X-ray observations of CR7 might then confirm the presence of the DCBH. However, this limit is already obtained with 180 ks of integration time on Chandra, meaning that a stringent test might only be possible with the next generation of X-ray telescopes.

Updated photometric observations of CR7 by Bowler et al. (2016) seem to suggest that CR7 is a more standard system, maybe a low-mass, narrow-line AGN or a young, low-metallicity starburst with the presence of binaries. Nonetheless, Pacucci et al. (2017) and Agarwal et al. (2017) have proposed a DCBH model that is consistent with the new IRAC-2 photometry within 1$\sigma$ and 3$\sigma$, respectively.

Pacucci et al. (2016) expanded this study, by looking for a method to select DCBH candidates in deep multi-wavelength surveys. Picking three infrared photometric filters (H band, IRAC-1 and IRAC-2), they found that the infrared colors of DCBH are much redder than the ones of other classes of high-$z$ sources (see Fig. 4). This redness should be caused by the very large column densities of the halos hosting these sources, which should be able to decrease the energy of outgoing photons. The authors selected two DCBH candidates in the CANDELS/GOODS-S field, at photometric redshifts $z \sim 6$ and $z \sim 10$ (see Giallongo et al. 2015). These sources are also detected by Chandra\footnote{It is worth noting that the detection of X-ray emission in these sources as well as their redshift is still controversial (e.g. Georgakakis et al. 2015; Kim et al. 2015; Cappelluti et al. 2016).}, suggesting that some kind of high-energy process is well underway in their locations. These sources represent the best candidates discovered so far for being the first DCBH ever observed.
\section{Dust emission in $z\sim 6$ quasars}
\begin{figure*}[t!]
\centering
\includegraphics[angle=0,scale=.23]{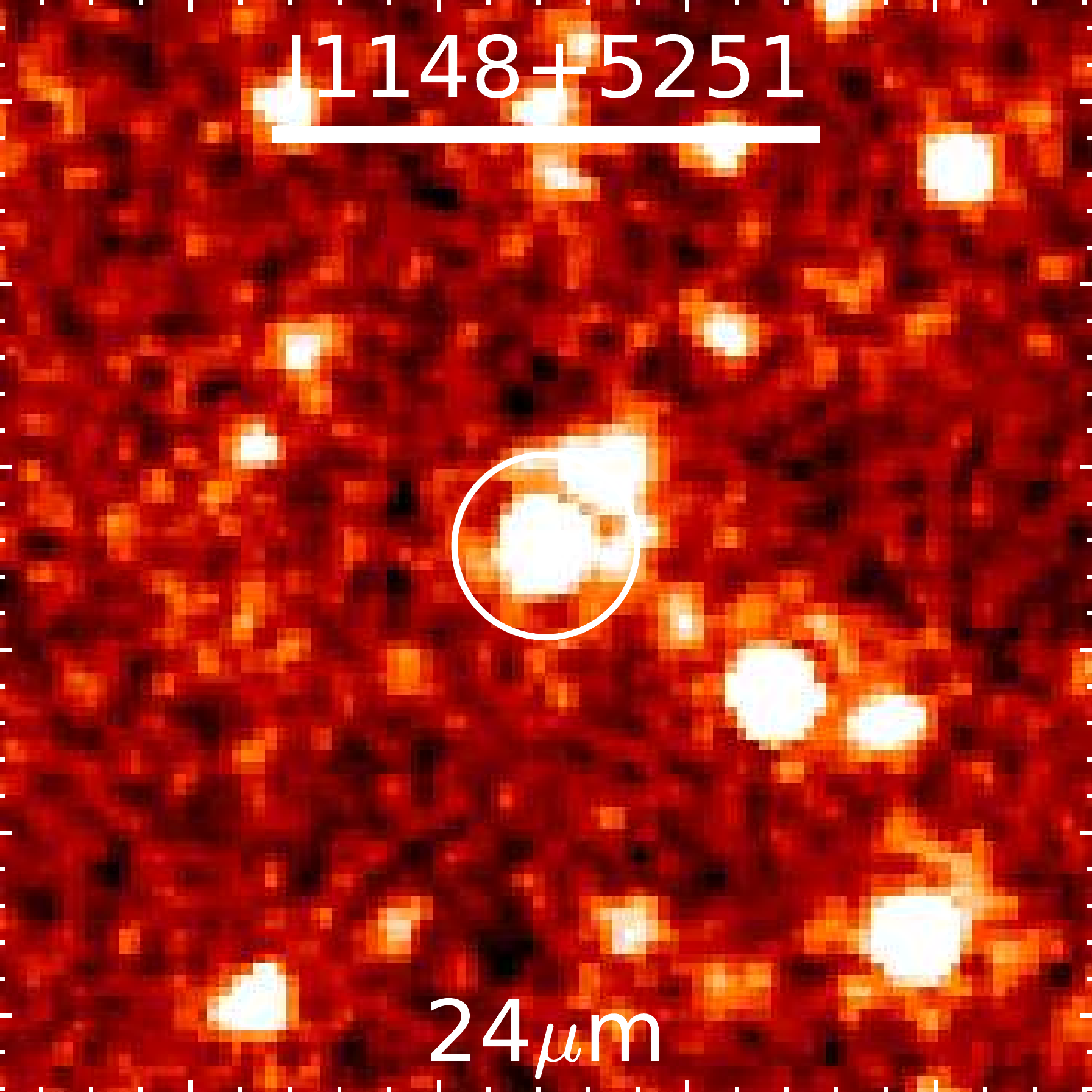}
\includegraphics[angle=0,scale=.23]{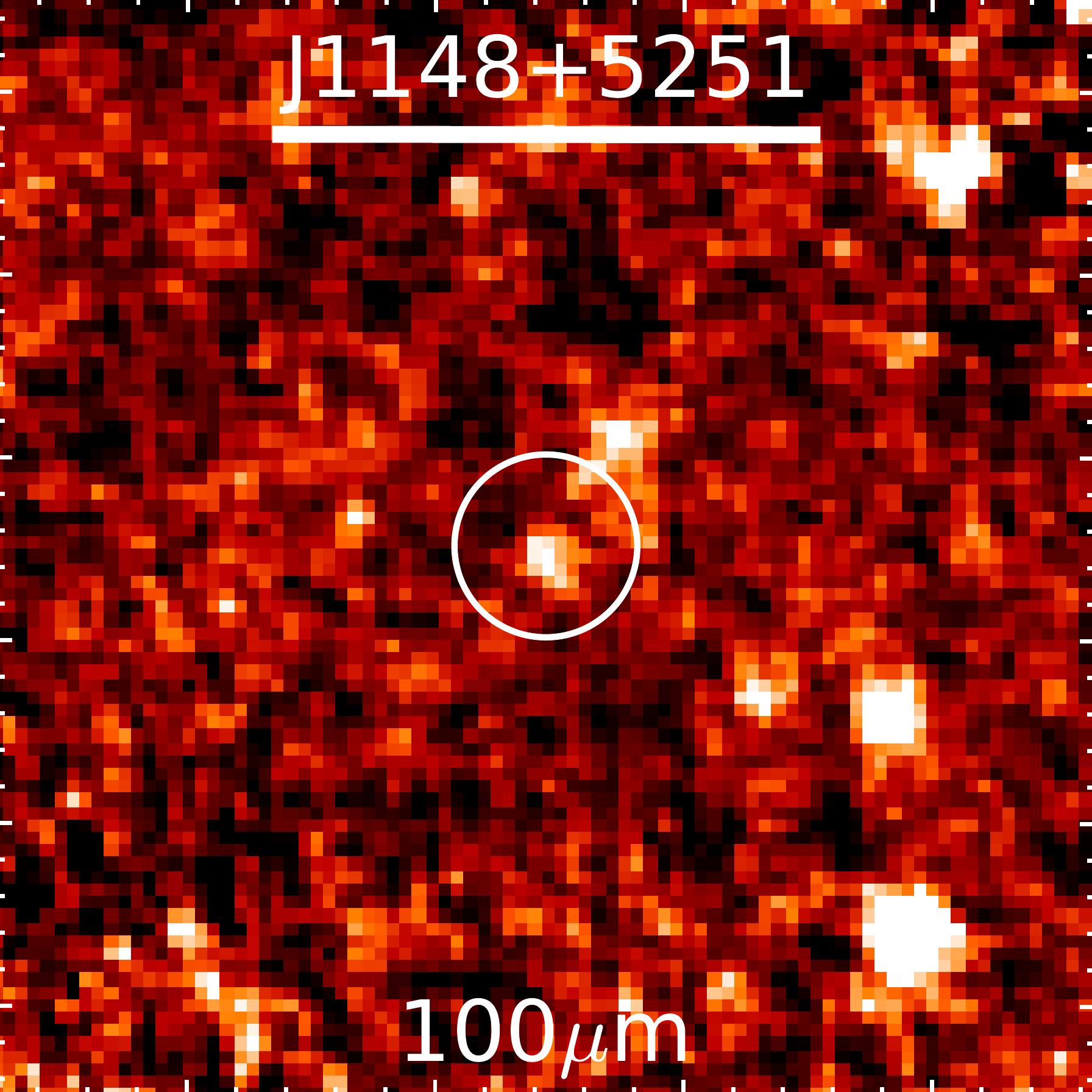}
\includegraphics[angle=0,scale=.23]{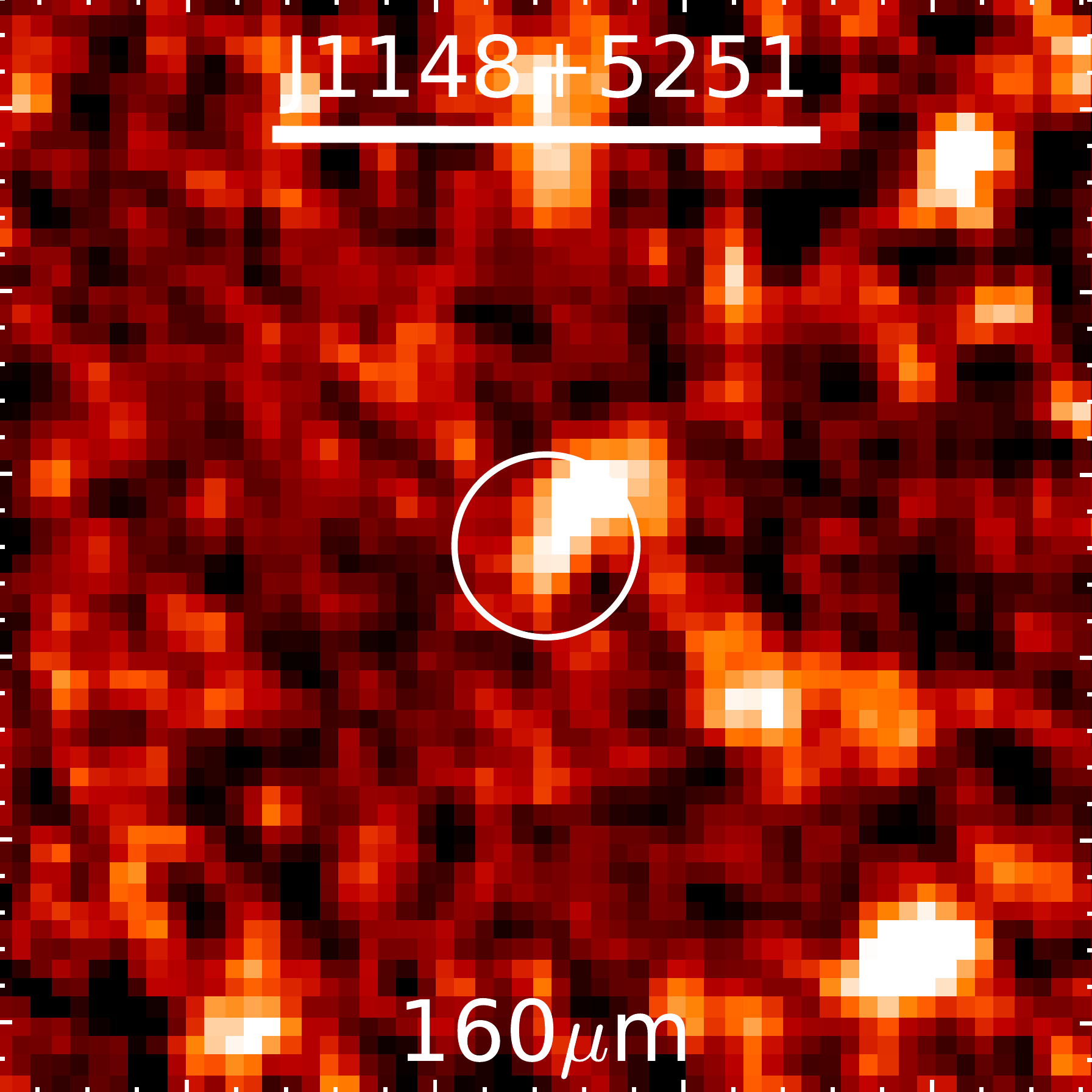}
\includegraphics[angle=0,scale=.23]{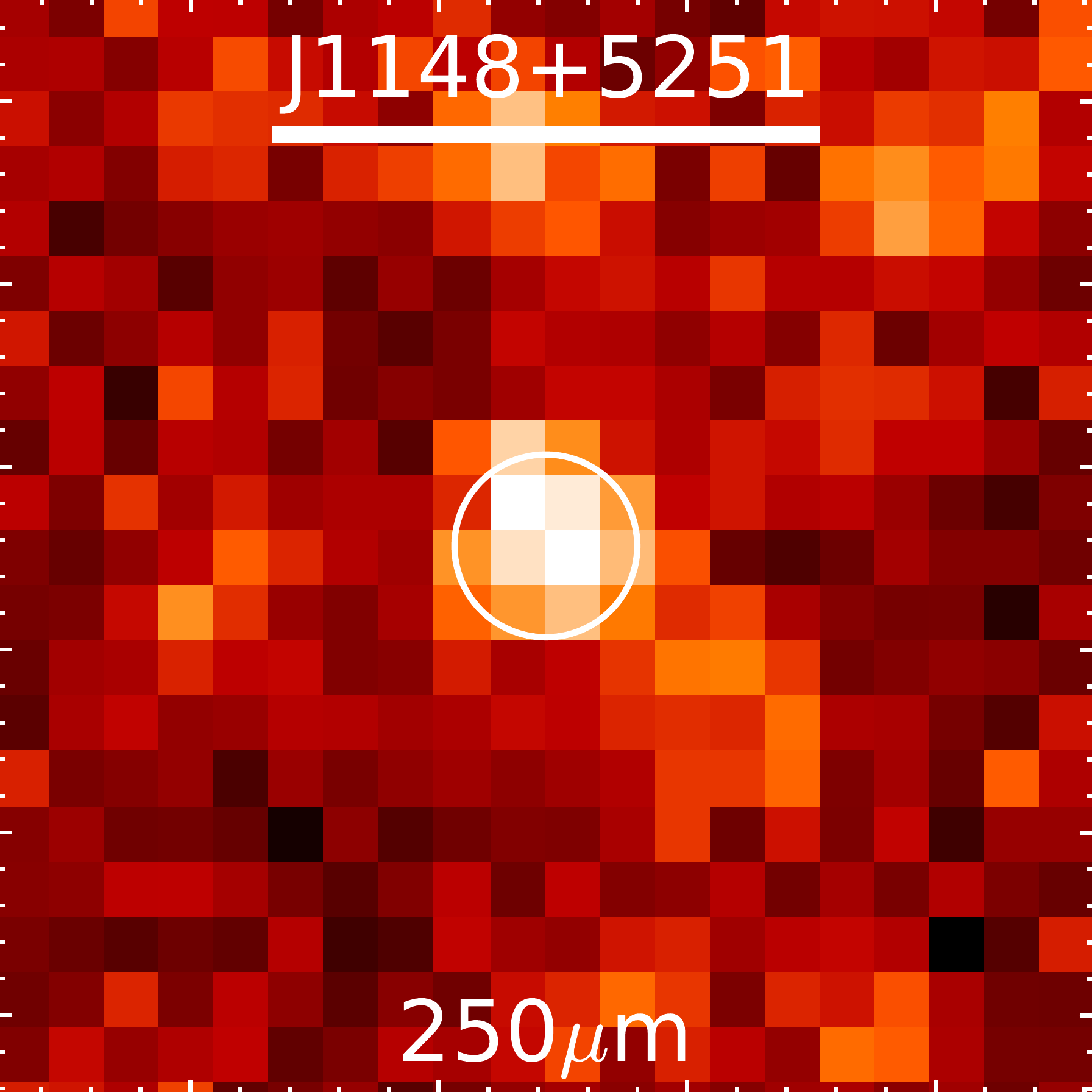}
\caption{Maps of J1148 at 24~$\mu$m (obtained with MIPS), 100 and 160 $\mu$m (PACS), and 250$\mu$m (SPIRE) from left to right. All images are 2 arcmin on a side and North is to the top with East to the left. The circle indicating the position of the quasar has a diameter of 20 arcsec. PACS observations reveal the presence of a secondary object $\sim 10^{"}$ north-west of the quasar. A possible counterpart is also seen in the MIPS map. The source complex is also detected in the SPIRE band, but the spatial resolution is too low to identify a possible double source. Adapted from Fig. 1 of Leipski et al. (2013) and reproduced by permission of the AAS.}
\label{herschel}
\end{figure*}
\begin{figure}
\includegraphics[width=8.9cm]{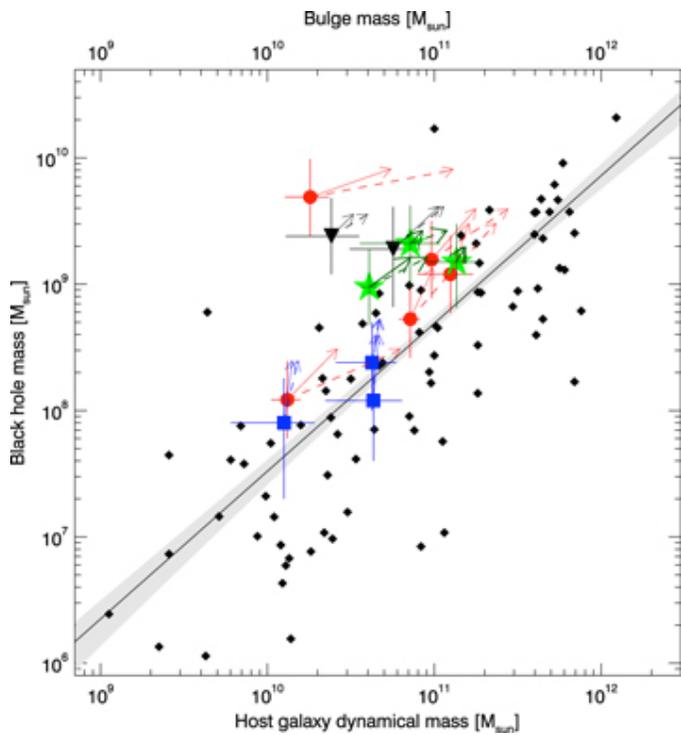}
\caption{The black hole mass vs. dynamical mass relation of $z \sim 6$ quasars. The black diamonds are values obtained for local galaxies (taken from Kormendy \& Ho 2013). The solid line and the shaded area shows the local $M_{BH}$ vs. $M_{bulge}$ relation derived by Kormendy \& Ho (2013). Reproduced from Fig. 12 of Venemans et al. (2016) by permission of the authors and the AAS.}
\label{fig:VenemansMsigma}
\end{figure}
\begin{figure}
\includegraphics[width=8.9cm]{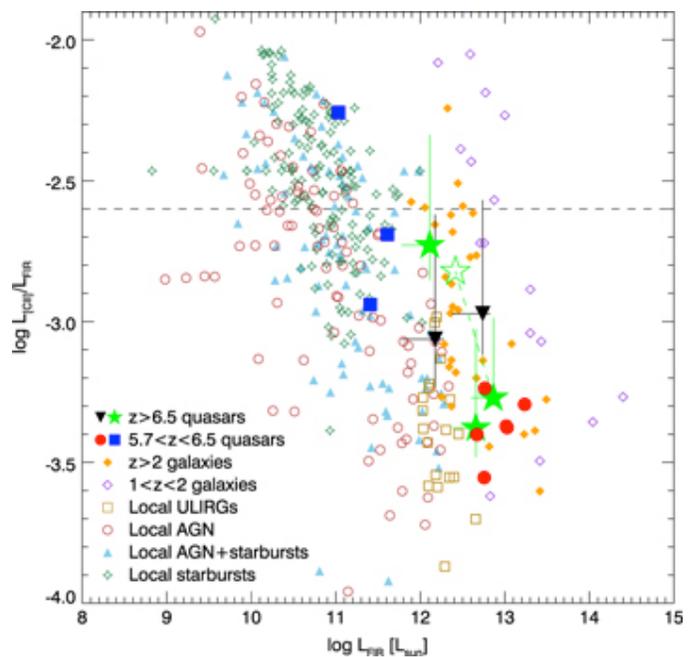}
\caption{The relation between FIR luminosity and the ratio of [CII] to FIR luminosity at different redshift. The horizontal dashed line show the average value of the [CII] to FIR ratio found in local star-forming galaxies. Reproduced from Fig. 5 of Venemans et al. (2016) by permission of the authors and the AAS.}
\label{fig:VenemansCII_FIR_ratio}
\end{figure}
\begin{figure*}
\includegraphics[height=1.5in]{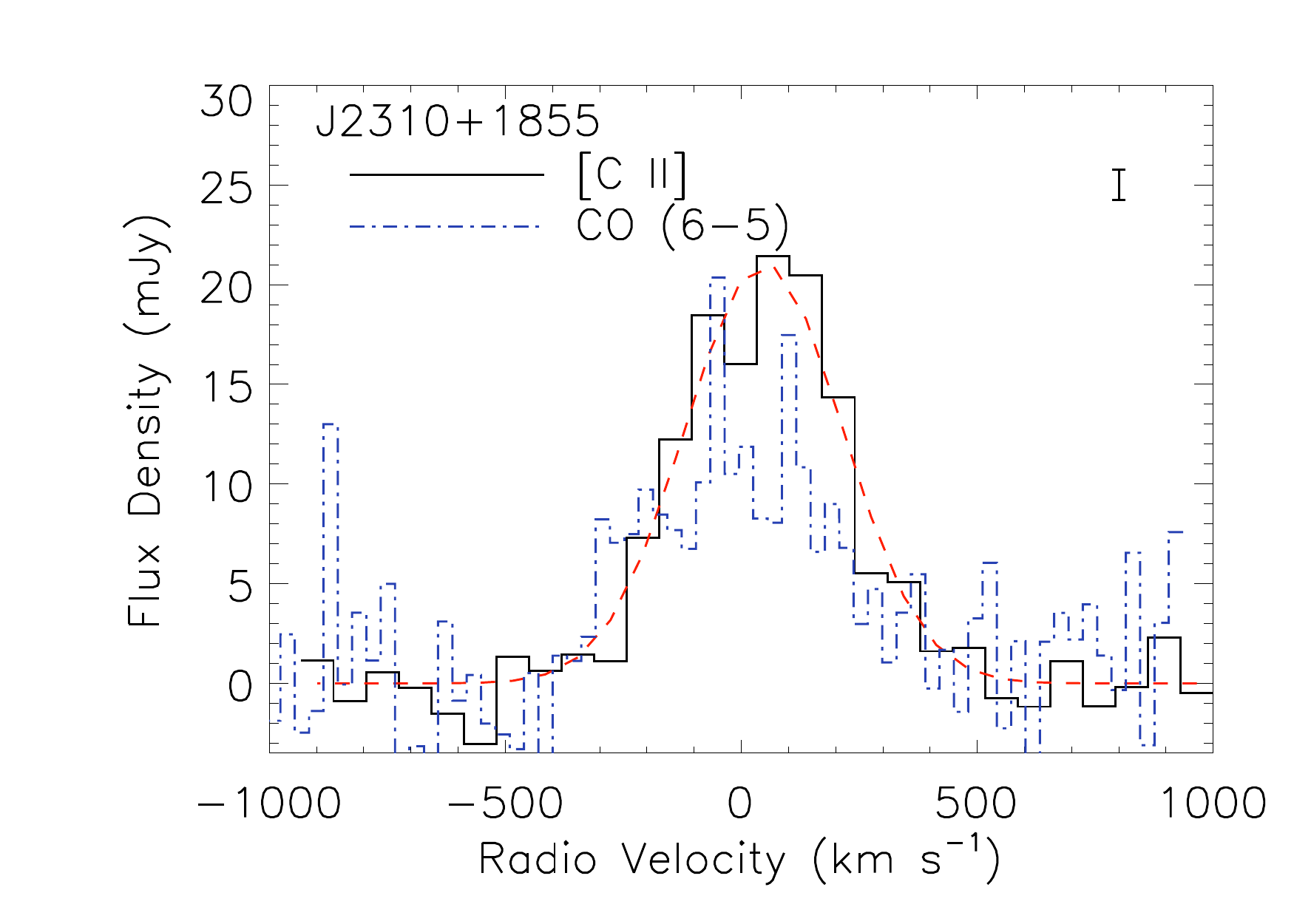}
\hspace*{0.2in}
\includegraphics[height=1.75in]{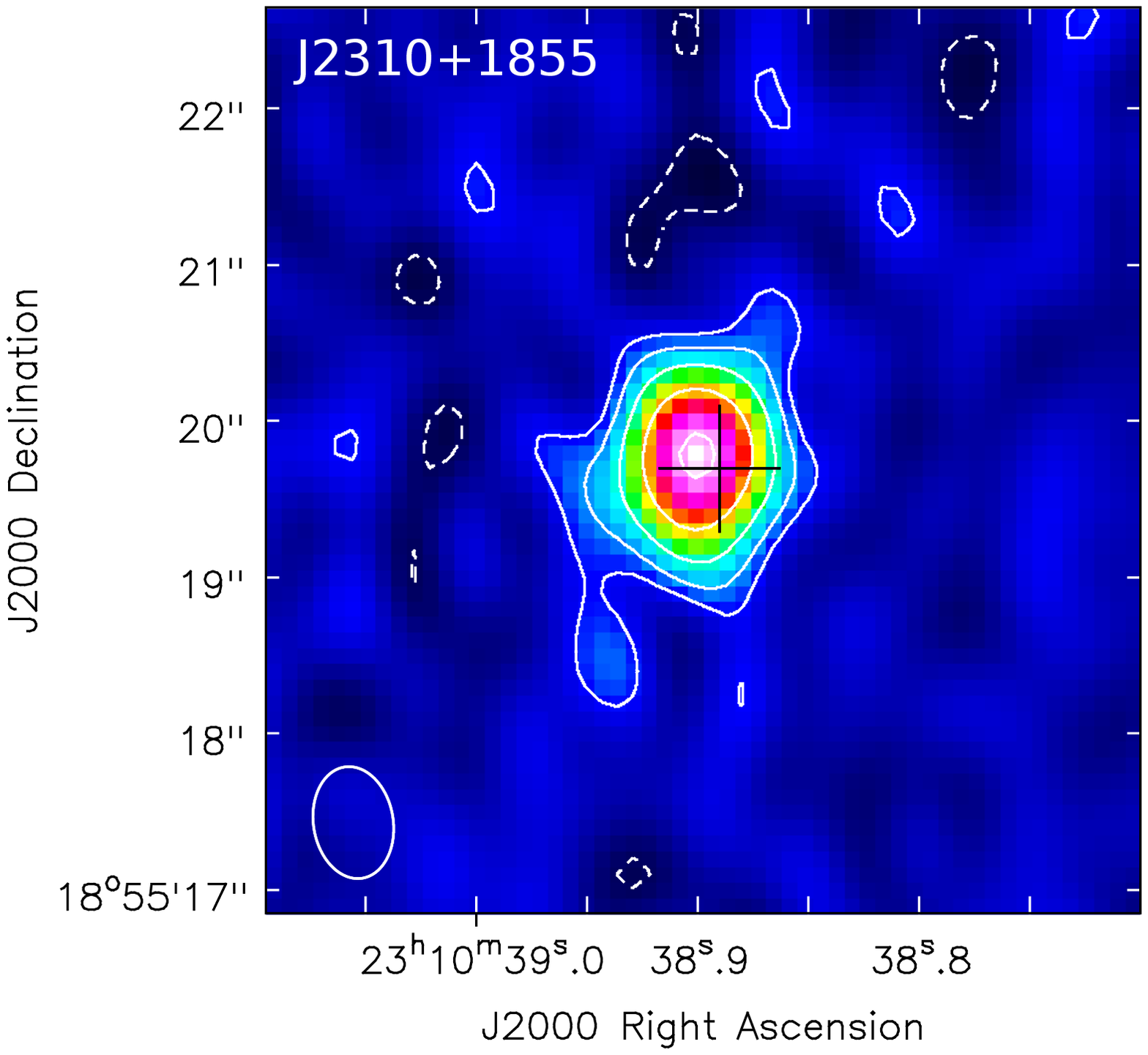}
\includegraphics[height=1.75in]{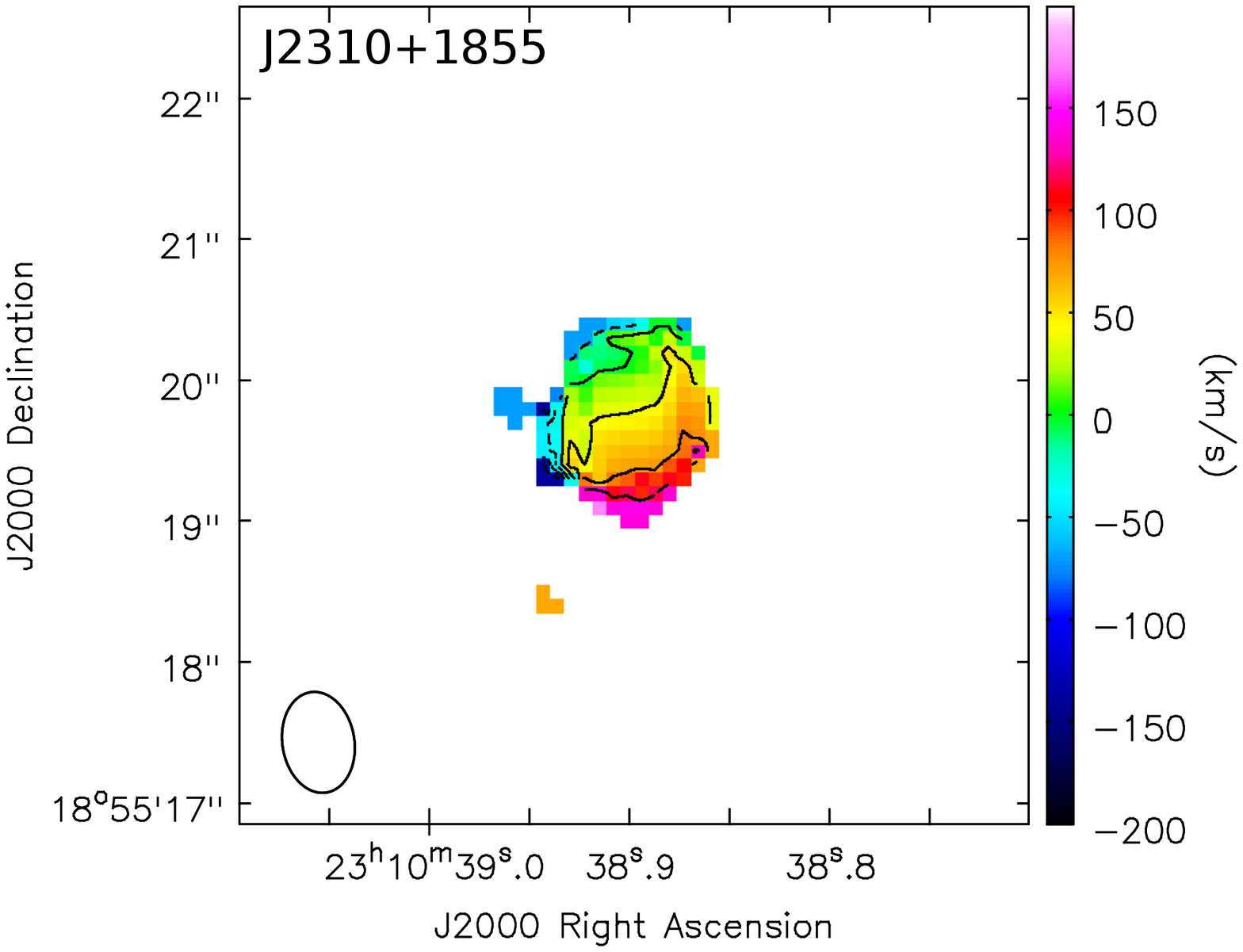}\\
\includegraphics[height=1.5in]{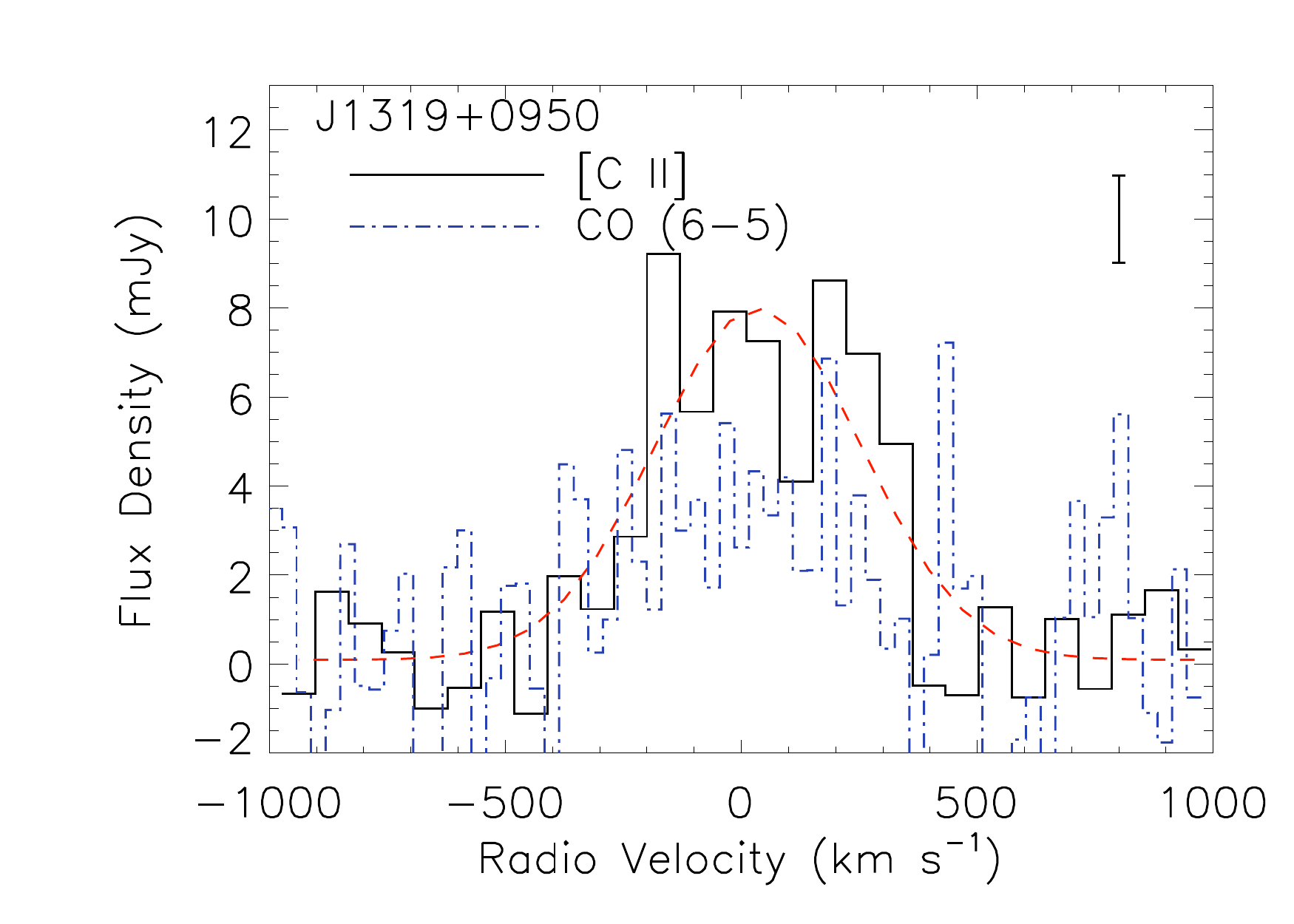}
\hspace*{0.22in}
\includegraphics[height=1.75in]{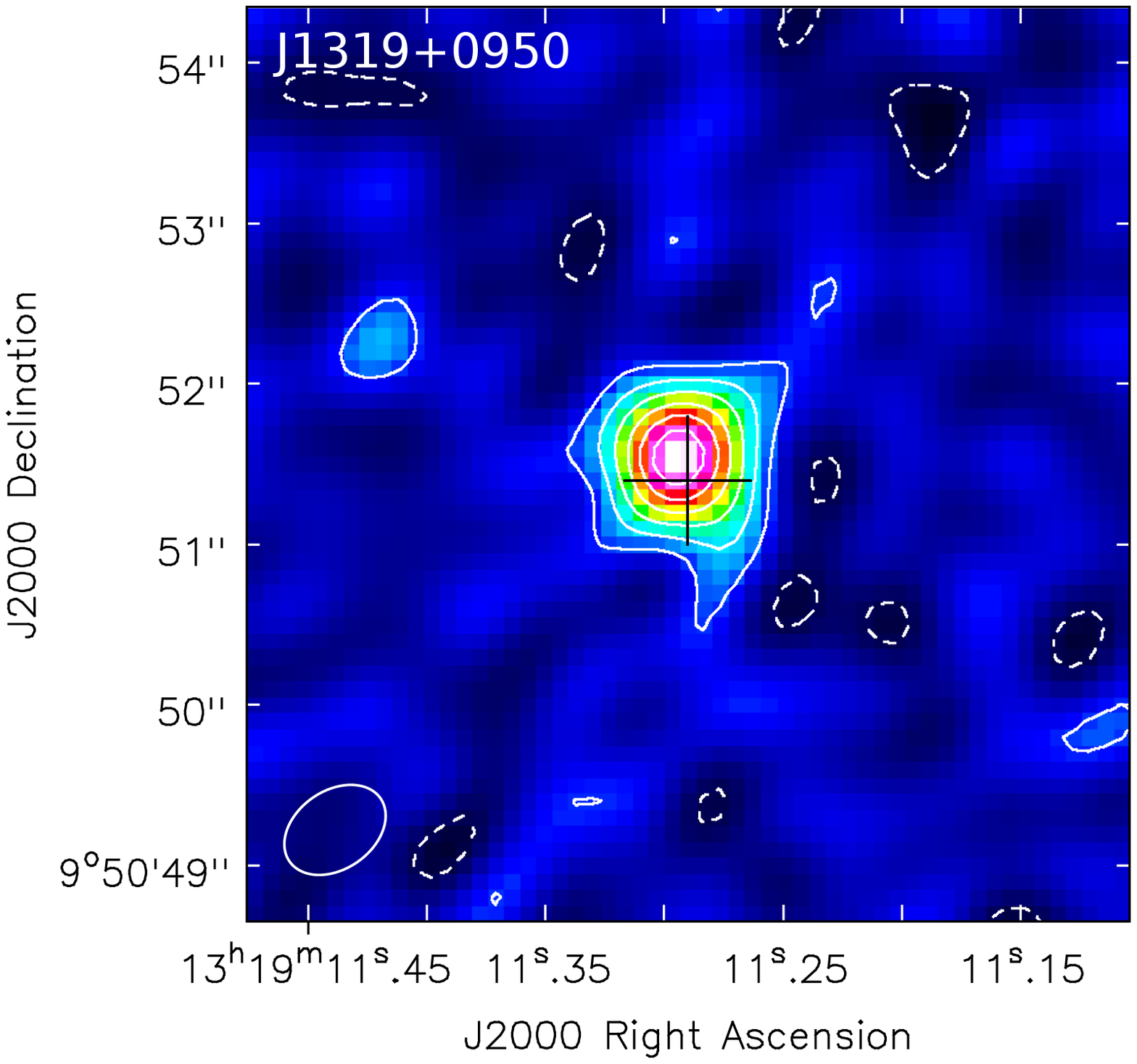}
\hspace*{0.05in}
\includegraphics[height=1.75in]{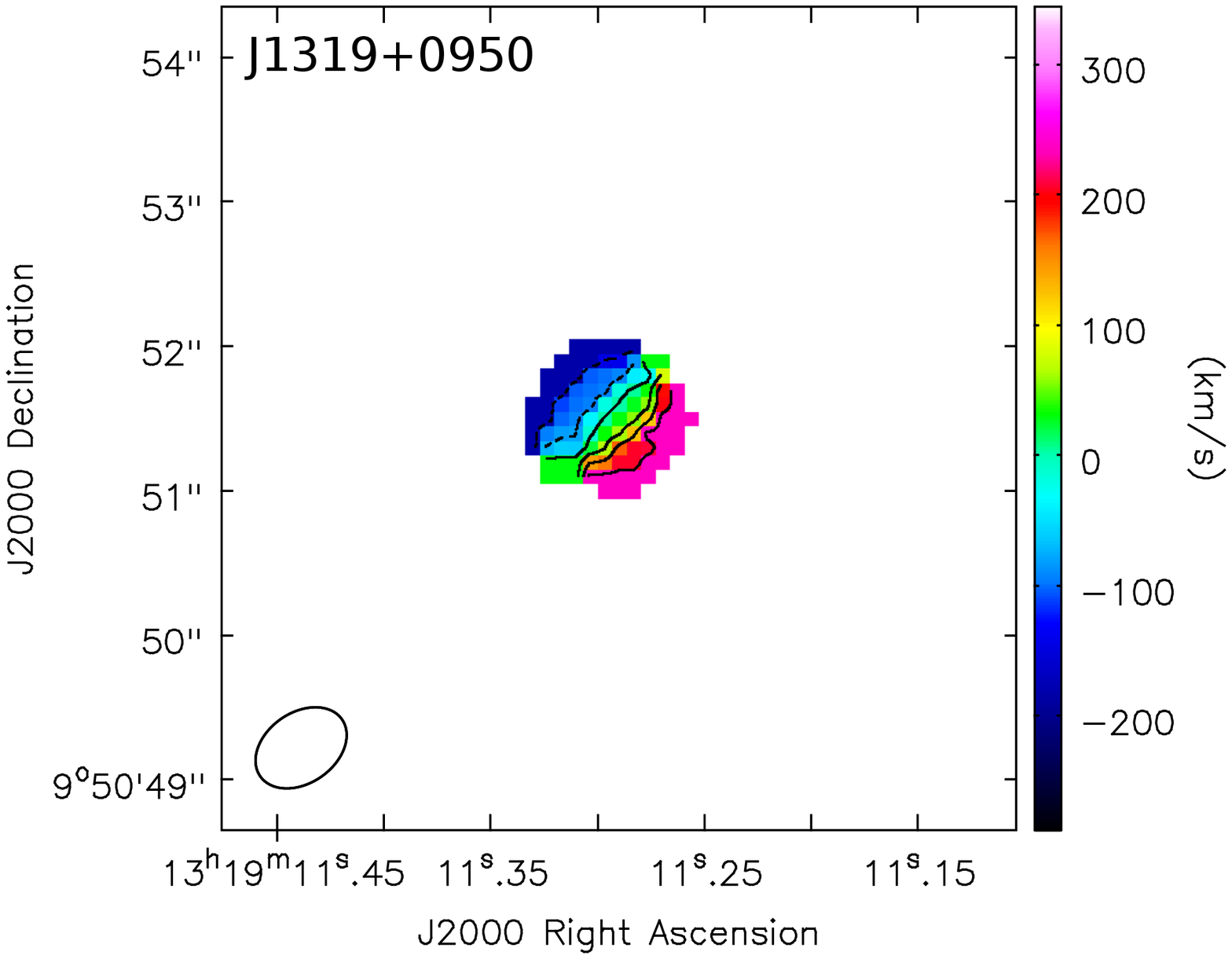}\\
\includegraphics[height=1.5in]{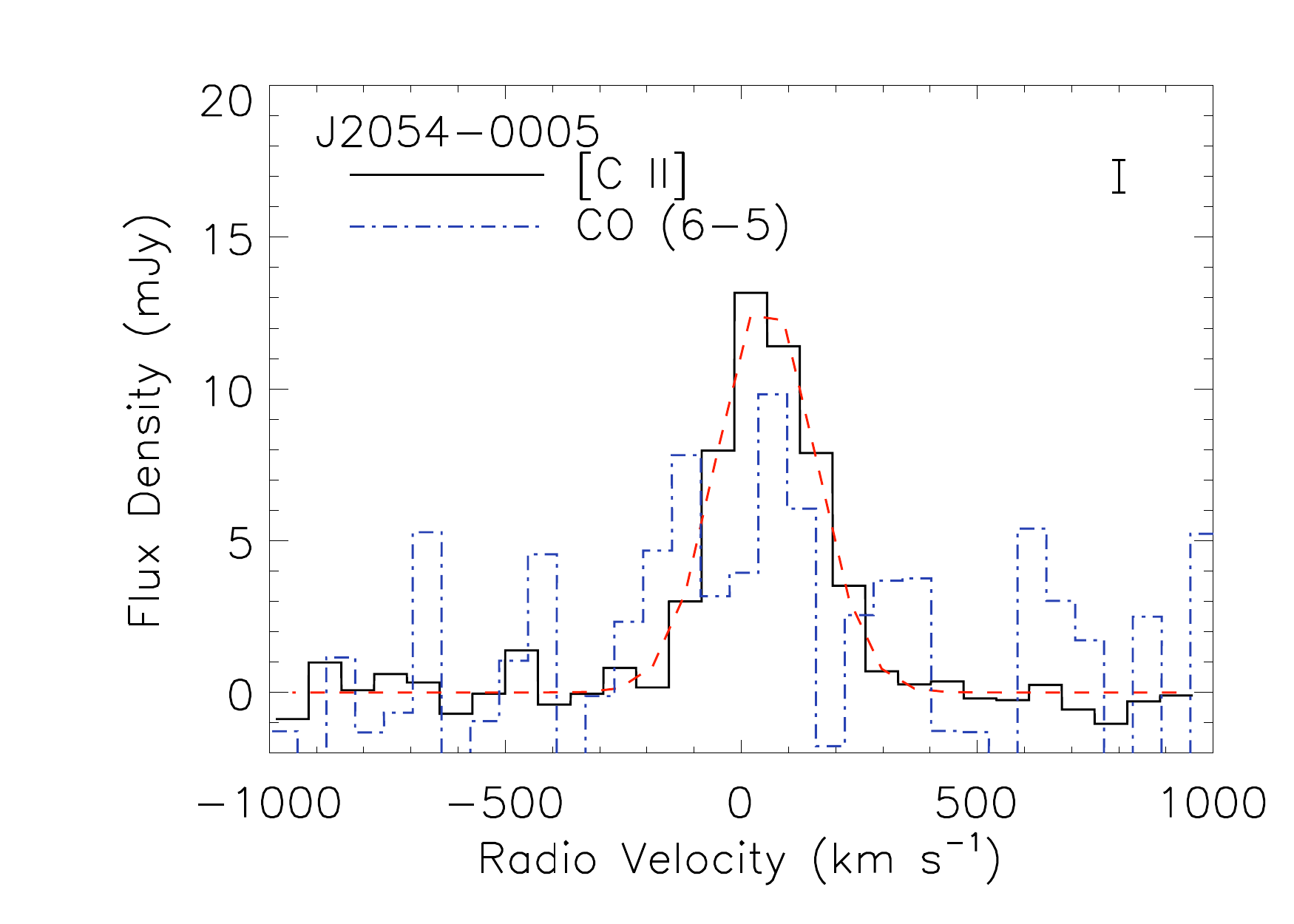}
\hspace*{0.15in}
\includegraphics[height=1.75in]{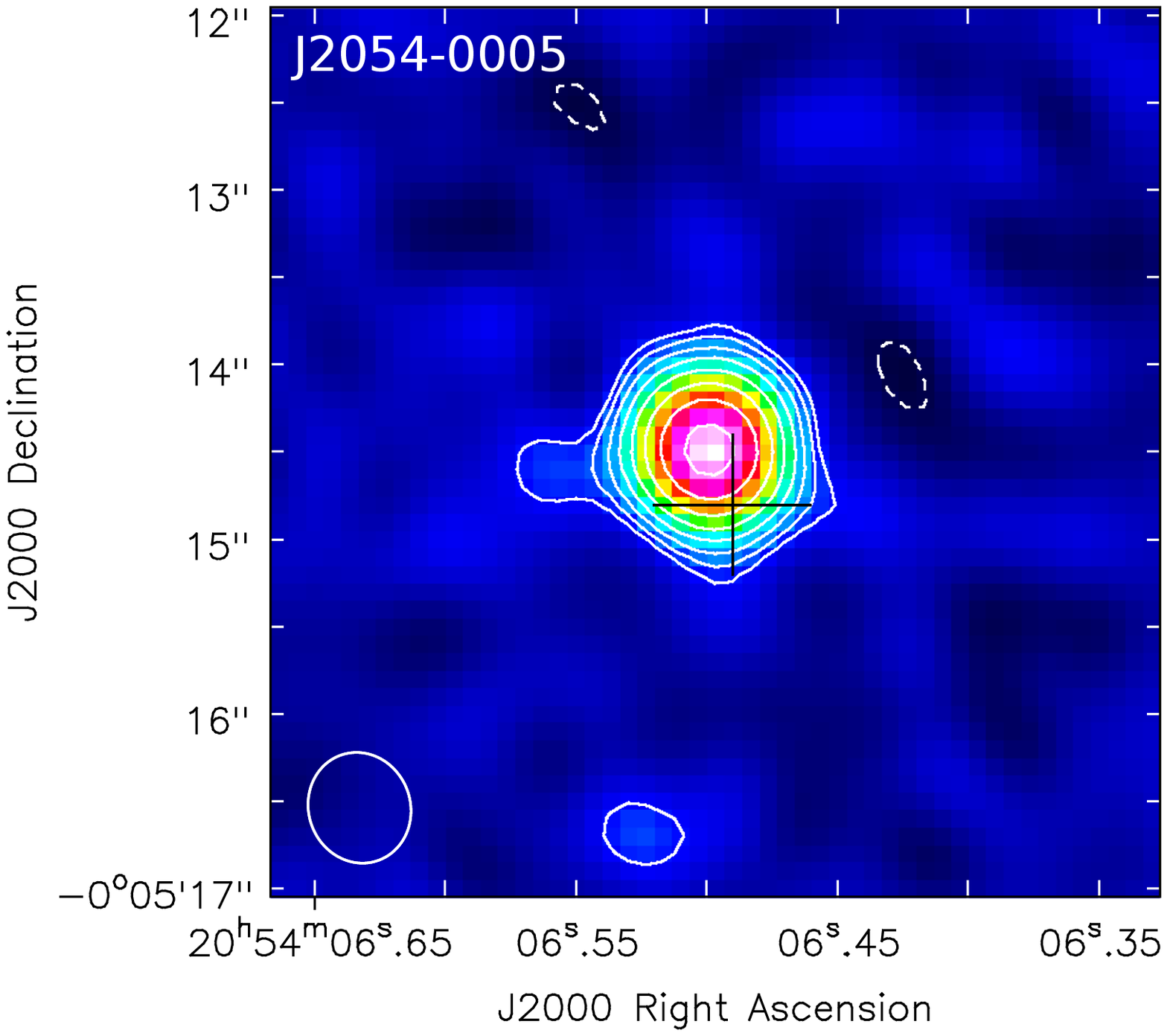}
\hspace*{0.02in}
\includegraphics[height=1.75in]{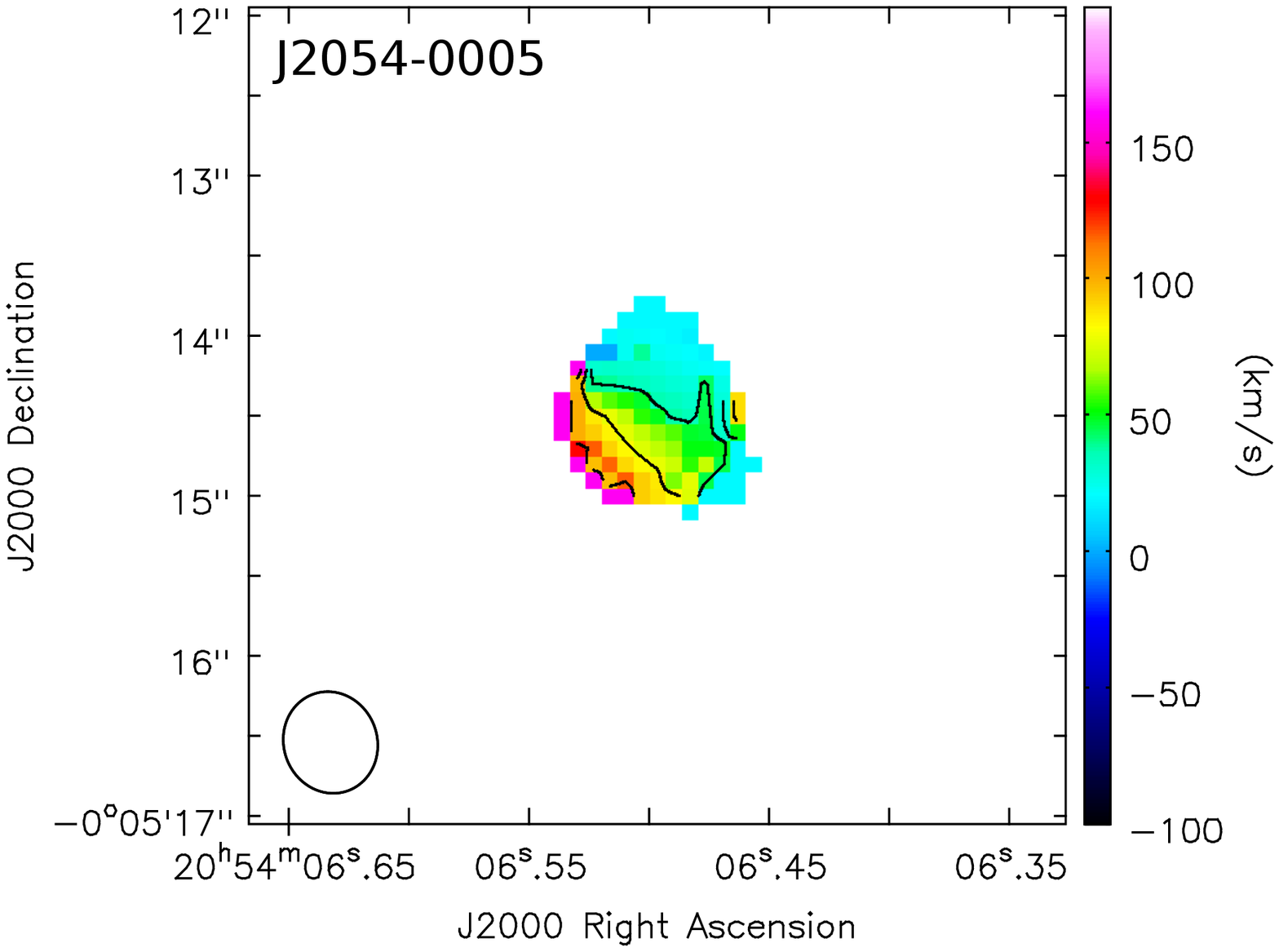}\\
\caption{{\bf Left panels}: [CII] line spectra (black solid line) of the five ALMA detected quasars, together with the previous CO (6-5) detections from PdBI (blue dotted line, scaled to the [CII] line) and a Gaussian fit to the [CII] line (red dashed line). {\bf Middle panels}: [CII] line velocity-integrated map. {\bf Right panels}: Line intensity-weighted velocity maps using pixels detected at $\geq 4\sigma$. Adapted from Fig. 2 and Fig. 5 of Wang et al. (2013) and reproduced by permission of the AAS. }
\label{fig:WangCIIspec}
\end{figure*}
Spectra of $z\sim 6$ quasars reveal strong metal emission lines
from the surrounding gas, with super-solar metallicity\footnote{We note that these metallicity measurements refer to the BLRs, small ($\lesssim$ 1 pc), low mass ($10^2-10^4~$M$_{\odot}$) nuclear regions that may be characterized by larger metallicity values with respect to the ones in the host galaxy. For example, Lyu et al. (2016) have recently shown that the infrared SEDs of $z\gtrsim 5$ quasars can be reproduced by combining a low-metallicity ($Z\sim 0.3 Z_{\odot}$) galaxy template with a standard AGN template (see their discussion in Sec. 5.2).} and almost no evolution with redshift (e.g. Fan et al. 2004, Juarez et al. 2009).
This suggests vigorous recent star formation in their host galaxies which enriched quasar environment.
Luminous high-redshift quasars are likely located in the densest environments in the early 
Universe. However, direct imaging of the stellar light of host galaxies is extremely  difficult for luminous high-redshift quasars. 
For example, Mechtley et al. (2012) carried out deep HST/WFC3 near-infrared observations of quasar SDSS J1148+5251 ($z=6.42$, hereafter J1148), attempting to detect the UV stellar continuum from young stars in the quasar host galaxy. After careful PSF subtraction, it yields a non-detection, constraining a UV-based star formation rate to be less than 250 $M_{\odot}$ yr$^{-1}$. 
Similarly, Decarli et al. (2012) used HST/WFC3 narrow-band imaging to search for extended Ly$\alpha$ nebular emission powered by star formation in the host galaxies of two luminous $z\sim 6$ quasars; the non-detections put a similar upper limit on the unobscured star formation rate in those systems. 

On the other hand, in observed FIR to mm wavelengths, the radiation is dominated by the reprocessed radiation from cool/warm dust in the host galaxy. Assuming the dust is heated by star formation, similar to the process common in star forming galaxies at local universe and at high redshift, the FIR to mm observations would provide  the best window to study host galaxy properties while minimizing contamination from the quasar light. 

In the pre-ALMA era, such observations have been obtained with MAMBO, Herschel, CSO, SCUBA, Spitzer, and are limited to relatively bright sources, with continuum sensitivity around 1 mJy at observed 1 mm wavelength, corresponding to $\sim 150 \, \mu$ in the rest-frame at $z \sim 6$.
Wang et al. (2007, 2008, 2011a) have led a systematic search of dust emission at 250 GHz (1.2 mm) over 30 quasars at $z \sim 6$, covering a wide luminous range, using the MAMBO bolometer on IRAM 30-meter telescope (see also Priddey et al. 2003, Bertoldi et al. 2003a). They found that about 1/3 of high redshift quasar host galaxies have luminosity comparable to those of 
hyper-luminous IR galaxies ($L_{FIR}= L(8-1000~\mu m) \sim 10^{13}$ L$_\odot$), with the fraction remaining roughly constant from low redshift to $z\sim 6$. The brightest objects in the sample are also detectable in shorter wavelength with Herschel, CSO and SCUBA, allowing a grey-body model fit to their SEDs. 
Figure \ref{fig:WangSED} shows examples of the $z\sim 6$ quasar SEDs; they can be well fit with warm/cool dust at $T\sim40-50$~K, consistent with the star formation providing the dust heating. This corresponds to an average dust mass of order $10^8$ M$_\odot$, and star formation rate of several hundred up to a few thousand $M_{\odot}$ yr$^{-1}$ for the mm bright quasars. Black hole accretion is indeed likely accompanied by intense star formation at the level expected to the formation of galactic bulge in massive galaxies, and the overall SEDs of the quasar hosts are consistent with that of Ultra-Luminous Infrared Galaxies (ULIRGs) such as Arp 220, with most of the star formation taken place in obscured phase, consistent with the HST rest-frame UV constraints. 

Wang et al. (2011a) also surveyed a sample of faint $z\sim 6$ quasars using IRAM/MAMBO (see also Omont et al. 2013), and found that although many are not detected with high significance at 1~mJy level, stacking of all undetected sources produces a strong signal, with average flux level of $\sim 0.5$~mJy, that still corresponds to a high star formation rate of a few hundred $M_{\odot}$ yr$^{-1}$. Figure \ref{fig:WangFIR} summarizes their results of the MAMBO survey, in comparison with low redshift sample. Overall, the FIR luminosity follows a weak dependence on the quasar bolometric luminosity (dominated by optical/UV accretion energy): $L_{FIR} \sim L_{bol}^{0.6}$. The overall FIR activity level is higher at $z\sim 6$ compared to local quasars at similar bolometric luminosity, suggesting a higher level of star formation activity accompanying the growth of massive early black holes. 

Complete infrared SEDs of $z>5$ quasars, obtained with Herschel and Spitzer (e.g. Leipski et al. 2010, 2013), have been used to disentangle the star formation versus AGN contribution to the total FIR emission ($8<\lambda_{RF}<1000~\mu$m). Leipski et al. (2013) have fitted the observed SED with four distinct components: i) a power law in the UV/optical regime to account for the accretion disk emission; ii) a blackbody peaking in the restframe NIR to account for hot dust emission ($T\sim 1300$~K); iii) a clumpy torus model to account for MIR emission arising from the AGN heated ``dusty torus''; iv) a modified black body to account for FIR emission powered by star formation. The result of this study is that star formation may contribute 25-60\% to the bolometric FIR luminosity, with strong variations from source to source. In Fig. \ref{herschel}, we show the results obtained in the case of J1148. The figure shows maps at 24~$\mu$m (obtained with MIPS), 100 and 160 $\mu$m (PACS), and 250~$\mu$m (SPIRE) from left to right. In this specific case, $\sim 50$\% of the FIR emission is due to dust heated by star formation, and $\sim 50$\% is due to the AGN heated torus.
\begin{figure}
\vspace{-2cm}
\includegraphics[clip=true, trim=3.5cm 4.7cm .2cm 7.2cm,angle=90,width=1.\columnwidth]{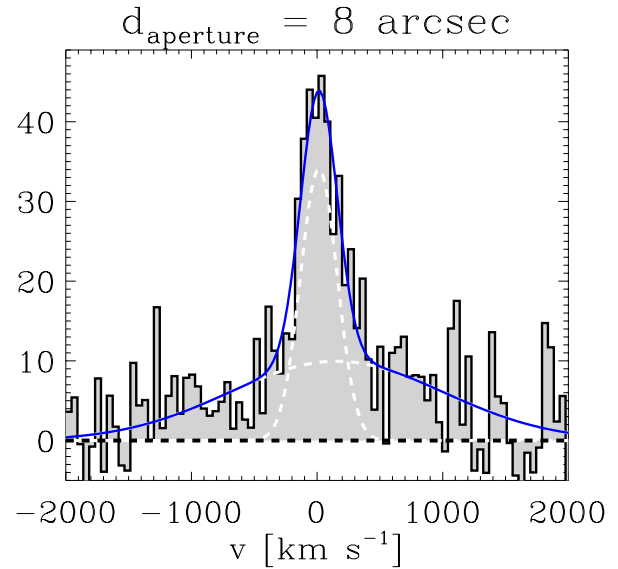}\quad
\caption{IRAM PdBI continuum-subtracted spectrum of the [CII] 158$\mu$m emission line of J1148. The spectrum shown has been extracted using a circular aperture with a diameter of 8 arcsec. For display purposes, the spectrum has been re-binned into channels of 46.8 \kms. The Gaussian fits to the line profiles (using a narrow and a broad Gaussian component to fit respectively the narrow core, tracing quiescent gas, and the broad wings, tracing the outflow) are performed on the original non-binned spectrum (channels of 23.4 \kms). Adapted from Fig. 1 of Cicone et al. (2015). Credit: Cicone Claudia, A\&A, 574, 14, 2015, reproduced with permission \copyright ~ESO.}
\label{fig:cii_spec_z6}
\end{figure}



\begin{figure*}
\includegraphics[angle=270,width=8.5cm]{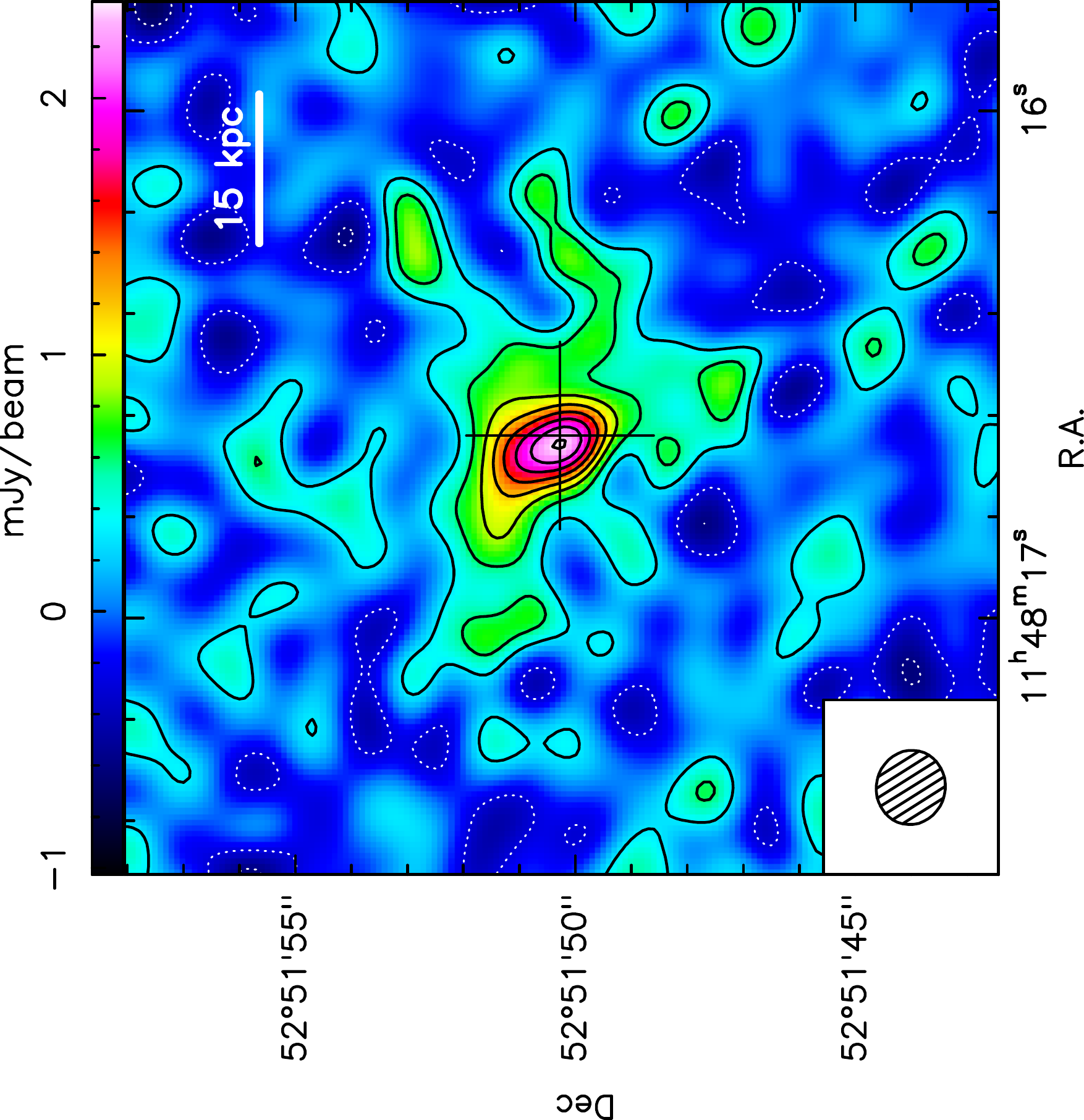}
\includegraphics[angle=270,width=8.5cm]{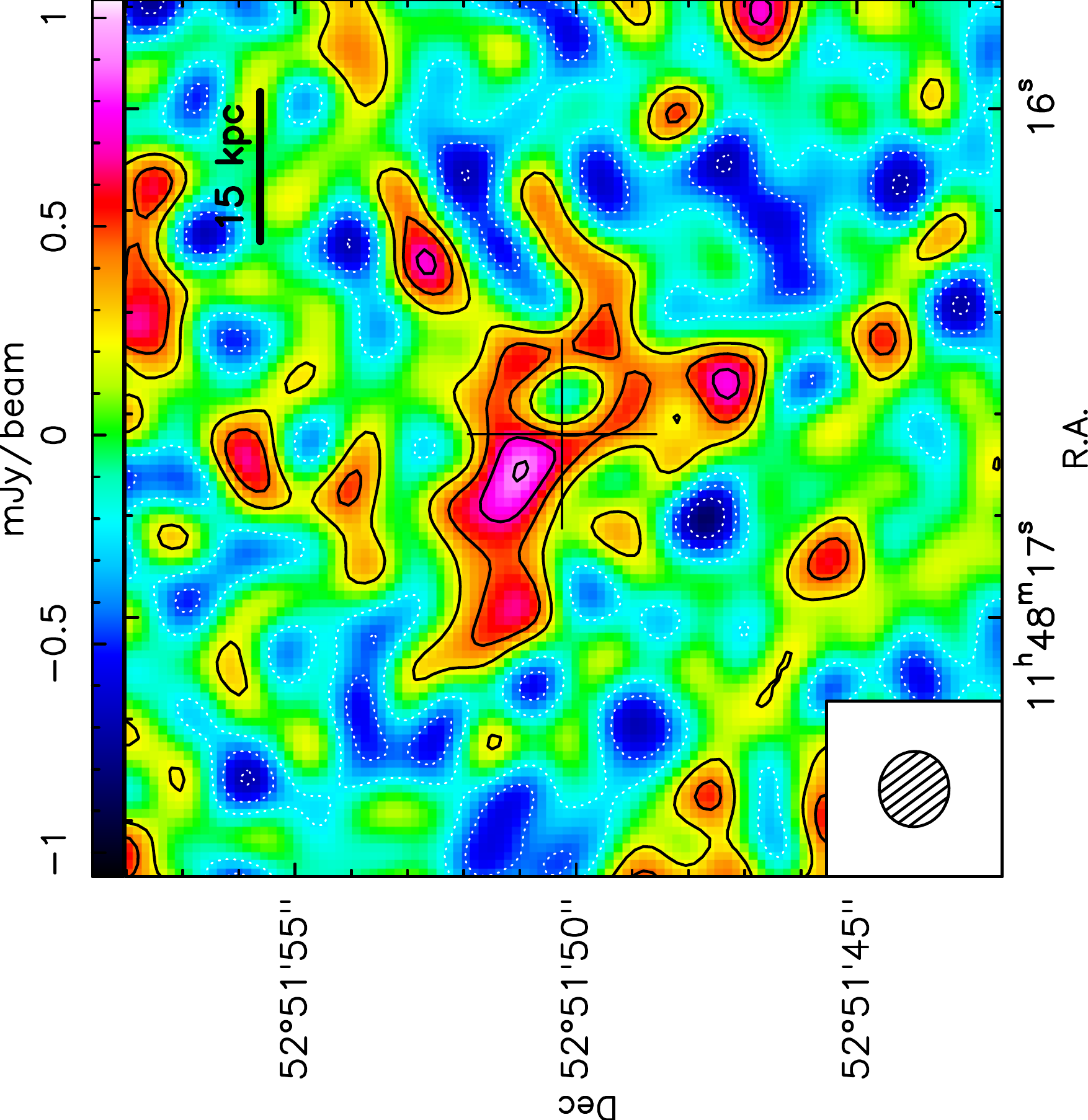}
\caption{{\bf Left:} IRAM PdBI continuum-subtracted map of the total [CII] 158$\mu$m emission of J1148, integrated within $v\in(-1400, 1200)$ \kms. Negative and positive contours are in steps of 3$\sigma$ (1$\sigma$ rms noise is 0.26 Jy beam$^{-1}$ km s$^{-1}$). The synthesized beam (1.3''$\times$1.2'') is shown in the bottom left corner of the map. The cross indicates the pointing and phase centre, corresponding to the optical position of the quasar. Adapted from Fig. 2 of Cicone et al. (2015). {\bf Right:} [CII] map obtained by integrating only the high velocity wings, within [-1400,-300]~km/s and within
[400,1200]~km/s, revealing the outflow extension. Adapted from Fig. 5 of Cicone et al. (2015). Credit: Cicone Claudia, A\&A, 574, 14, 2015, reproduced with permission \copyright ~ESO.}
\label{fig:cii_map_z6}
\end{figure*}

Observations of Wang et al. (2011a), Omont et al. (2013), and Leipski et al. (2013) are at the limit of previous generation facilities. The majority of quasars at high-redshift are still undetected individually,  with potential strong selection effects from the low S/N data and non-detections, as well as possible luminosity dependent biases. ALMA has revolutionized the field of submm and mm studies of galaxies. Even in its early science observation phase, ALMA have represented a factor of  $> 10$ increase in observing efficiency for both continuum and line detections at submm wavelengths comparing to current facilities.

Venemans et al. (2016) presented sensitive ALMA observations of three quasars at $6.6<z<6.9$ in both dust continuum and [CII] emission. These observations reached a rms level of 0.1~mJy on the continuum. All three objects are clearly detected with star formation rate of 100 to 1600 $M_{\odot}$ yr$^{-1}$ based on similar assumptions to Wang et al. (2011a). Willott et al. (2015) carried out the deepest yet ALMA observations of high-redshift quasars, targeting two $z\sim 6$ quasars at moderate luminosity and with black hole masses an order of magnitude lower than the average in the Wang et al. (2011a) sample. These observations reached a rms level on 0.03~mJy on the continuum, and showed a significantly lower FIR luminosity: the 1.2mm continuum luminosity is 0.05 - 0.2~mJy, corresponding to a FIR-based star formation rate of only 19 - 73 $M_{\odot}$ yr$^{-1}$. Clearly, the survey at the most luminous end does not tell the whole story of the black hole - starburst connection at high redshift, and we expect that comprehensive ALMA surveys in the coming years will unveil a complete picture of star formation around supermassive black holes at high redshift.

One assumption of all studies discussed above is that FIR dust emission in quasar hosts is powered by star formation similar to what observed in ULIRGs. As shown in Figures \ref{fig:WangSED}, this is consistent with the shape of quasar SED over wide wavelength range. Lyu et al. (2016) fit a family of SED models that consist of AGN and starburst templates to a sample of Spitzer+Herschel+mm SEDs (Leipski et al. 2014) finding a dominant starburst contributions in the FIR bands probed by MAMBO and ALMA observations. On the theoretical side, Schneider et al. (2015) carried out the first detailed radiative transfer model to fit the broad-band multi-wavelength SED of the well observed quasar J1148. They found that the central AGN heating could contribute 30-70\% of the observed FIR emission, but starburst contribution is still consistent with a star formation rate of $\sim 1000~M_{\odot}$ yr$^{-1}$. 
\section{[CII] emission in $z\sim 6$ quasars}
\begin{table*}
\caption{Black hole masses ($M_{\rm BH}$), [CII] luminosities ($L_{\rm [CII]}$),  [CII] line widths (FWHM$_{\rm [CII]}$) of $z\sim 6$ quasars}
\label{table_obs}
\begin{center}
\begin{tabular*}{41.5pc}{@{}l\x c\x c\x c\x c\x c\x c@{}}
\hline\hline
\vspace {.2cm}
Name& [CII] redshift&$M_{\rm BH}~(M_{\odot})$&$L_{\rm [CII]}(L_{\odot})$& FWHM$_{\rm [CII]}$~(km~$s^{-1}$)&Ref$^a$\\
\hline
\vspace {.2cm}
J1120+0641$^b$&7.0842$\pm$0.0004&$2.4^{+0.2}_{-0.2}\times10^9$&1.2$\pm$0.2$\times10^9$&235$\pm$35&[1][2]\\
\vspace {.2cm}
J2348-3054&6.9018$\pm$0.0007&$2.1^{+0.5}_{-0.5}\times10^9$&1.9$\pm$0.3$\times10^9$&405$\pm$69&[1][3]\\
\vspace {.2cm}
J0109-3047&6.7909$\pm$0.0004&$1.5^{+0.4}_{-0.4}\times10^9$&2.4$\pm$0.2$\times10^9$&340$\pm$36&[1][3]\\
\vspace {.2cm}
J0305-3150&6.6145$\pm$0.0001&$9.5^{+0.8}_{-0.7}\times10^8$&3.9$\pm$0.2$\times10^9$&255$\pm$12&[1][3]\\
\vspace {.2cm}
J036+03&6.5412$\pm$0.0018&$1.9^{+1.1}_{-0.8}\times10^9$&5.8$\pm$0.7$\times10^9$&360$\pm$50&[4][5]\\
\vspace {.2cm}
J0210-0456&6.4323$\pm$0.0005&$0.8^{+1.0}_{-0.6}\times10^8$&3.0$\pm$0.4$\times10^8$&189$\pm$18&[6][7]\\
\vspace {.2cm}
J0100+2802&6.3258$\pm$0.0010&$1.2^{+0.2}_{-0.2}\times10^{10}$&3.6$\pm$0.5$\times10^9$&300$\pm$77&[8][9]\\
\vspace {.2cm}
J2229+1457&6.1517$\pm$0.0005&$1.2^{+1.4}_{-0.8}\times10^8$&6.0$\pm$0.8$\times10^8$&351$\pm$39&[6][10]\\
\vspace {.2cm}
J1319+0950&6.1330$\pm$0.0007&$2.1^{+3.8}_{-1.4}\times10^9$&4.4$\pm$0.9$\times10^9$&515$\pm$81&[10][11]\\
\vspace {.2cm}
J2054-0005&6.0391$\pm$0.0001&$0.9^{+1.6}_{-0.6}\times10^9$&3.3$\pm$0.5$\times10^9$&243$\pm$10&[10][11]\\
\vspace {.2cm}
J0055+0146&6.0060$\pm$0.0008&$2.4^{+2.6}_{-1.4}\times10^8$&8.3$\pm$1.3$\times10^8$&359$\pm$45&[6][10]\\
\vspace {.2cm}
J2310+1855&6.0031$\pm$0.0002&$2.8^{+5.1}_{-1.8}\times10^9$&8.7$\pm$1.4$\times10^9$&393$\pm$21&[10][11]\\
\vspace {.2cm}
J1044-0125&5.7847$\pm$0.0007&$1.1^{+1.9}_{-0.7}\times10^{10}$&1.6$\pm$0.4$\times10^9$&420$\pm$80&[10][11]\\
\vspace {.2cm}
J0129-0035&5.7787$\pm$0.0001&$1.7^{+3.1}_{-1.1}\times10^8$&1.8$\pm$0.3$\times10^9$&194$\pm$12&[10][11]\\
\hline\hline
 \end{tabular*}
\end{center}
\tabnote{$^a$ First and second labels represent references for $M_{\rm BH}$ and [CII] observations, respectively. $^b$ See Venemens et al. (2017) for follow-up observations. References: [1] De Rosa et al. (2014); [2] Venemans et al. (2012); [3] Venemans et al. (2016); [4] Venemans et al. (2015b); [5] Ba$\rm \tilde{n}$ados et al. (2015); [6] Willott et al. (2010); [7] Willott et al. (2013); [8] Wu et al. (2015); [9] Wang et al. (2016); [10] Willott et al. (2015); [11] Wang et al. (2013)}
\end{table*}
Atomic fine structure line emission, and in particular the [CII] 158 $\mu$m line, is thought to be the dominant coolant of the ISM, tracing the cold neutral medium and photon-dominated regions associated with star formation.
The [CII] line is the strongest ISM line in the local universe. However, because of its wavelength in the FIR, for long time it has been impossible to observe; thus, the first [CII] detections have been limited to low-redshift galaxies from space observations (see for example the recent works by Herrera-Camus et al. 2015 and Samsonian et al. 2016).\\ 
The first [CII] observations at high-redshift came from luminous quasars (e.g. Maiolino et al. 2005; Walter et al. 2006), since [CII] becomes observable from the ground at $z>4-5$ (e.g. Iono et al. 2006, Wagg et al. 2010, De Breuck et al. 2011, Wagg et al. 2012, Cox et al. 2011). This line is important for understanding the high-redshift universe because of its high luminosity, its connection to star formation, and the ability to acquire detailed kinematics using high spatial resolution radio interferometric observations. In particular, [CII] observations in $z\sim 4-5$ quasars have shown the presence of close ($<$50 kpc) massive star-forming galaxies interacting with the quasar hosts, possibly indicating that mergers play some role in the early, fast growth of SMBHs (e.g. Gallerani et al. 2012, Trakhtenbrot et al. 2017b).

The [CII] line is detected in quasars up to the highest redshift at $z=7.1$ (Venemans et al. 2012; Venemans et al. 2017). 
However, before ALMA, only a handful of high-redshift quasars are detected in [CII] (e.g. Ba$\rm \tilde{n}$ados et al. 2015; Wang et al. 2016) and only J1148 has sufficiently high S/N and resolution for detailed analysis of its host galaxy ISM properties (Walter et al. 2009, Maiolino et al. 2012).

Three early ALMA results highlight the power of using [CII] as a probe of star formation in high-redshift quasar host galaxies and the co-evolution of supermassive black holes and galaxies at early epoch.
The [CII] luminosity of $z\sim 6$ quasars observed with ALMA (Wang et al. 2013; Willott et al. 2015; Venemans et al. 2016) ranges from 0.6 to 8.7 $\times 10^9 L_{\odot}$. In all cases, [CII] lines are clearly resolved with FWHM ranging from 190 to 520 km s$^{-1}$. In Table \ref{table_obs}, we report all the detections of [CII] emission obtained so far in $z\sim 6$ quasars, along with their black hole masses. 

By combining the [CII] and FIR luminosity measurements it has been possible to compare the $L_{CII}/L_{FIR}$ ratio of $z\gtrsim 6$ quasars with results obtained in local galaxies (see Fig. \ref{fig:VenemansCII_FIR_ratio}). It results that the $L_{CII}/L_{FIR}$ ratio follows the low redshift trend of decreasing towards FIR brighter objects. The origins of this effect, generally called ``[CII] deficit'', are not definitively understood and it is still not clear whether it is driven by the presence of an X-ray source or simply by specific galaxy ISM properties. For instance, the central AGN may provide a substantial contribution to the dust heating and FIR emission, determining a lower $L_{CII}/L_{FIR}$ ratio in the nuclear region (e.g. Sargsyan et al. 2012) and/or its strong X-ray radiation may reduces the CII abundance (e.g. Langer \& Pineda 2015). Moreover, compact starbursts tend to show substantially smaller $L_{CII}/L_{FIR}$ ratios with respect to extended and more diffuse systems (e.g. D\'iaz-Santos et al. 2013). Finally, also the ISM metallicity can affect both [CII] and FIR emissions, though its final role on the ``[CII] deficit'' is controversial (e.g. De Breuck et al. 2011, Vallini et al. 2015). Given the large scatter of the observed $L_{CII}/L_{FIR}$ ratio at any given FIR luminosity, further observations are required to reach any conclusive understanding of the ``[CII] deficit'' origins (see further discussions in  Trakhtenbrot et al. 2017b, Narayanan \& Krumholz 2017, Brisbin et al. 2015, Graci\'a-Carpio et al. 2011, Luhman et al. 2003). 

In the left panels of Figure \ref{fig:WangCIIspec}, we show the [CII] line spectra of the Wang et al. (2013) sample, compared with their CO (6-5) observations from PdBI. In the middle panels, [CII] emission maps of the same sources are presented. The right panels show the velocity maps of the Wang et al. (2013) sample. In fact, an especially exciting aspect of ALMA [CII] observations is the ability of spatially resolve the host galaxy kinematics. In the bright sample of Wang et al. (2013), four of the five sources are spatially resolved under subarcsec resolution, with a deconvolved size of 2-4 kpc. For those that are resolved in [CII], a clear velocity gradient is evident across the image, consistent with a strong velocity shear due to either rotation or strong galaxy interaction.\\ 
We note that two of the $z>6.6$ quasars in the Venemans et al. (2016) sample are also resolved with similar size measurements. However, their velocity structures are not consistent with a flat rotation curve. In addition, one object shows a large (1700 km s$^{-1}$) velocity shift between [CII] redshift and the redshift of the MgII low-ionization line, usually used as standard quasar systematic redshift. This wide range of morphology points to a large diversity of quasar host galaxy properties, potentially linked to the different evolutionary stages of host galaxy assembly. For two fainter objects in Willott et al. (2015), one is resolved (with similar size) and the other marginally resolved due to low S/N. 

With constraints on both sizes and velocity line width, one can make crude estimates of the dynamical mass of the host galaxy ($1.3\times 10^{10}<M_{\rm dyn}/\rm M_{\odot}<1.2 \times 10^{11}$), and place these high redshift quasars onto the equivalent of $M_{BH}$-$\sigma$ or $M_{BH}$-$M_{dyn}$ relations established at low redshift. Fig. \ref{fig:VenemansMsigma} shows the $M_{BH}$-$M_{dyn}$ relation at $z\sim 6$: luminous high-redshift quasars systematically deviate from the local relation in the sense that their host mass is low compared to local galaxies of the same black hole mass. This seems to imply that at least for the luminous systems observed here, black hole growth precedes galaxy assembly and quasars with the largest black holes are not necessarily in the most massive galaxies in early epochs. However, there are a number of caveats in reaching any firm conclusion with regard to the evolution of black hole/galaxy mass relation, given the still limited quality of spatially resolved data, the potential complex gas kinematics and the luminosity bias when targeting a flux selected sample. In particular, estimates of the stellar mass in $z\gtrsim 5$ quasars by Lyu et al. (2016), based on the infrared SED fitting, provide BH-galaxy mass ratios consistent with the local relation. Moreover, Valiante et al. (2014) suggested that since observations of high-$z$ QSOs are sensitive to the innermost 2.5-3 kpc, they are simply missing the galaxy star formation located on larger scales. Interestingly, the fainter quasars with lower black hole masses observed in Willott et al. (2015) have similar [CII] line width and size constraints.

\section{Evidence for quasar feedback at $z\sim 6$}
Quasars are invoked by most models and numerical simulations to 
quench star formation in massive galaxies, through massive radiation-pressure driven outflows.
The resulting system, cleaned of cold gas, is expected to passively evolve into local
massive elliptical galaxies (e.g. Silk \& Rees 1998, Granato et al. 2004, Di Matteo et al. 2005,
King et al. 2011, Zubovas \& King 2012, Fabian 2015). This process is also thought to contribute
to the steeply declining mass function of galaxies at high masses, by preventing the overgrowth
of galaxies. By halting both star formation and black hole accretion this process
may also be responsible for the correlation between black hole and stellar mass in the host spheroid.

Evidence for quasar-driven outflows has been found in local systems, especially through the discovery of massive
and powerful winds traced by the cold phase of the ISM (Sturm et al. 2011, Feruglio et al. 2010, Cicone et al. 2014).
These examples are only local ``laboratories'' of quasar-driven outflows. However, in order to explain
the properties of local elliptical massive galaxies (which have an old stellar population) the bulk of the star
formation quenching
must occur at high redshift ($z\sim 2$). Evidence for massive outflows at high redshift is much more sparse.
Although quasar-driven outflows have been seen in $z\sim 2$ quasars, these are generally traced by the ionized
phase (generally through the [OIII] 5007~\AA~or Ly$\alpha$ lines; Cano-Diaz et al. 2012, Harrison et al. 2014, Brusa et al. 2015, Carniani
et al. 2015, Carniani et al. 2016, Swinbank et al. 2015),
which generally probes only a small fraction of the outflowing mass. Evidence of
molecular outflows has been found in a few targets at $z\sim 2.3$ (Weiss et al. 2012, George et al. 2012).
\begin{figure}
\hspace{-0.8cm}
\includegraphics[angle=0,width=9.cm]{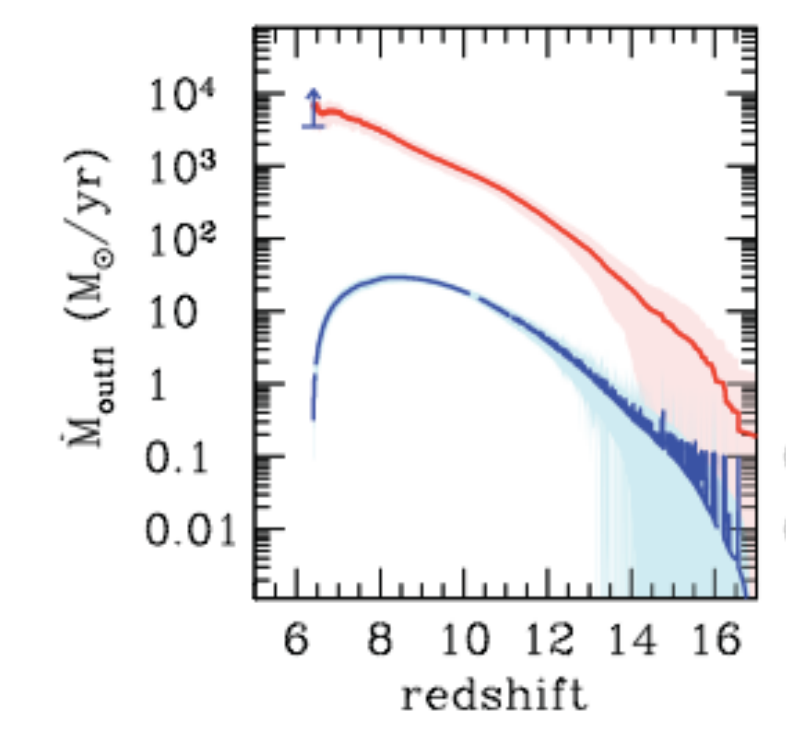}
\caption{Model developed by Valiante et al. (2012) showing the evolution of the quasar-driven (red) and SN-driven (blue)
outflow rates for a system with properties similar to J1148. The outflow rate observed in J1148 is indicated with a blue symbol as a lower limit, since the additional contribution to the outflow rate from other gas phases is not known. Adapted from Fig. 3 of Valiante et al. (2012) and reproduced by permission of the authors.}
\vspace{-0.5cm}
\label{fig:feedb_models_valiante}
\end{figure}
However, the discovery of passive and old (age 2-3 Gyr) galaxies at $z\sim 2.5$, when the age of the Universe was only
$\sim$3~Gyr (Kriek et al. 2006, Labbe et al. 2005, Saracco et al. 2005, Cimatti et al. 2004),
implies that quenching by quasar-driven outflows must have already been in place at $z\sim 6$, at least in a fraction
of very massive galaxies. Evidence for quasar-driven massive outflows has been recently found at $z>6$, in J1148 (Maiolino et al. 2012, Cicone et al. 2015). This outflow
has been identified through the detection of broad wings of the [CII] line (Fig. \ref{fig:cii_spec_z6}), tracing cold
gas expelled at velocities higher than 1000~km/s (see Table \ref{J1148}). The map of the [CII] emission extends over a region of about 30~kpc
(left panel of Fig. \ref{fig:cii_map_z6}). However, most of this extended emission is associated with the high velocity gas,
as shown in the right panel of Fig. \ref{fig:cii_map_z6}, implying that the outflow in this system is well developed and that cold gas has
been driven out to very large distances from the central active region.
The inferred outflow rate is very large ($1400\pm 300~\rm M_{\odot}~yr^{-1}$), comparable to or exceeding the
star formation rate. The implied depletion time of the gas in the host galaxy would be shorter
than 10 million years (assuming that there is no additional 
gas supply and that the outflow rate continues at this rate, but see additional discussion below on this regard),
implying that star formation would be quenched on very short timescales.
\begin{figure*}
\includegraphics[angle=0,width=17cm]{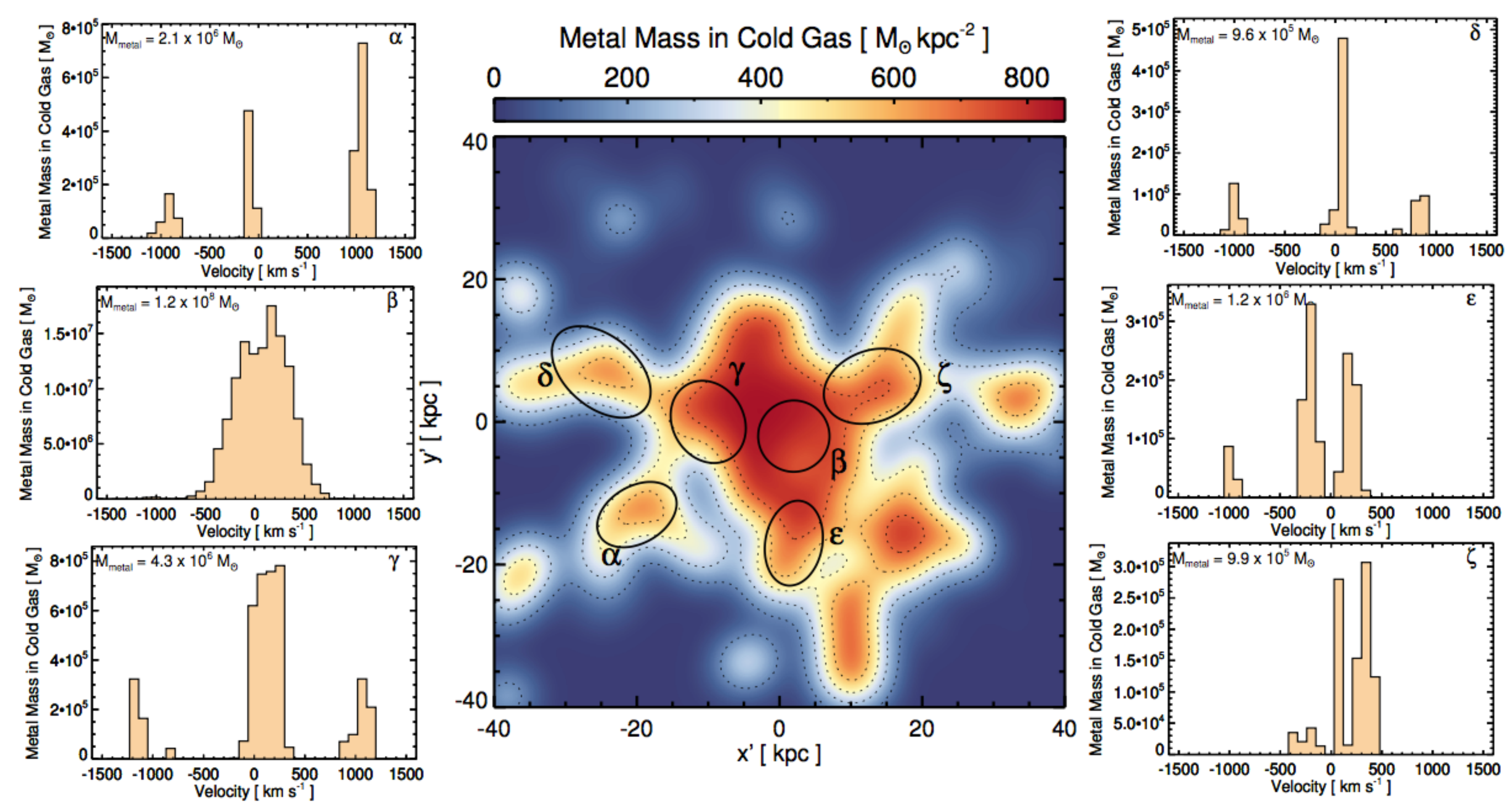}
\caption{Simulation of the cold gas distribution in a quasar host galaxy at z=6.4 (map in the middle). Figures on the left and right columns show the gas velocity distribution in some regions of the field of view, revealing that the outflowing gas can indeed reach velocities of about 1000 km/s. Reproduced from Fig. 3 of Costa et al. (2015).} 
\label{fig:feedb_models_costa}
\end{figure*}
\begin{figure*}
\includegraphics[angle=0,width=17cm]{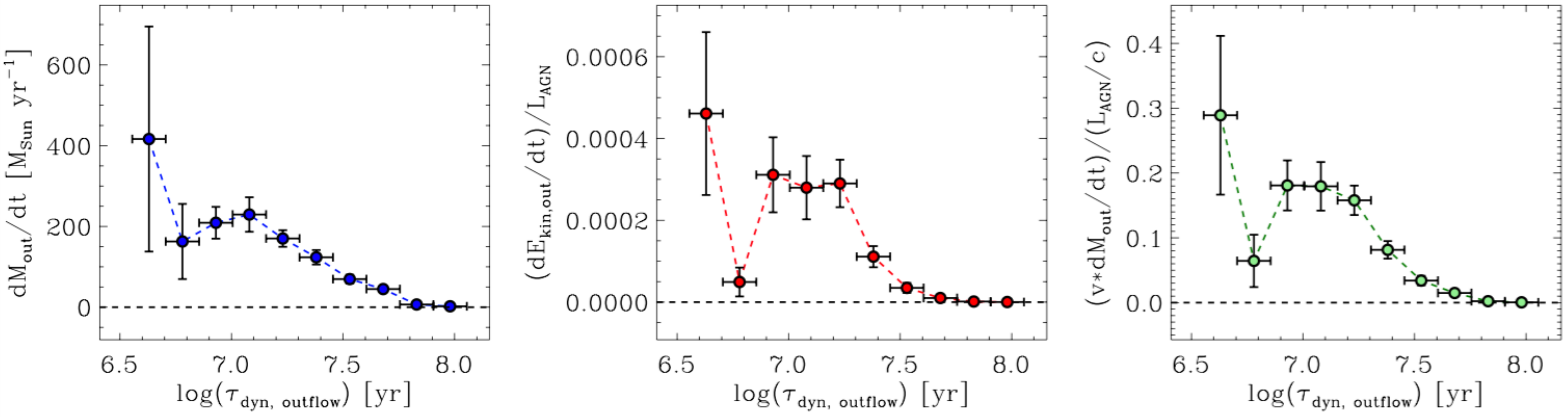}
\caption{Mass outflow rate (left), kinetic power (normalized to the AGN radiative luminosity, middle) and momentum rate (normalized
to $\rm L_{AGN}/c$, right) as a function of the dynamical time of the [CII] outflowing clumps, as observed in J1148. Reproduced from Fig. 7 of Cicone et al. (2015). Credit: Cicone Claudia, A\&A, 574, 14, 2015, reproduced with permission \copyright ~ESO.}
\label{fig:tdyn}
\end{figure*}
\begin{figure*}
\begin{center} 
\includegraphics[width=6.2cm]{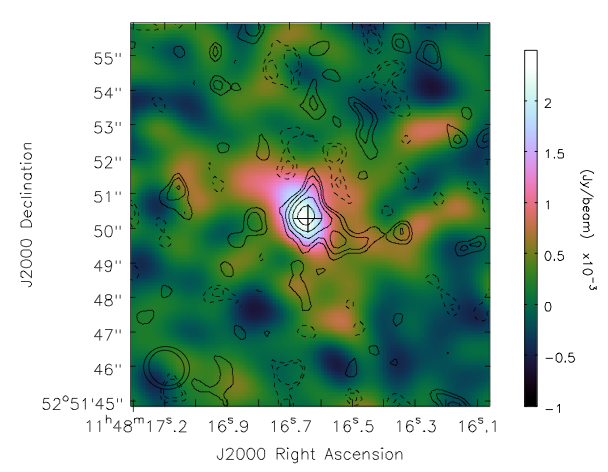}
\includegraphics[width=5.8cm]{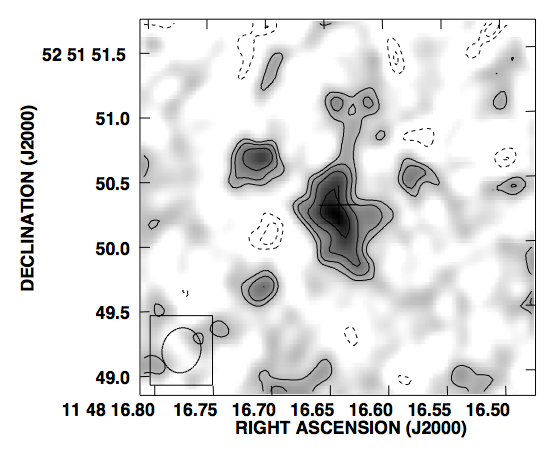}
\includegraphics[width=5.1cm]{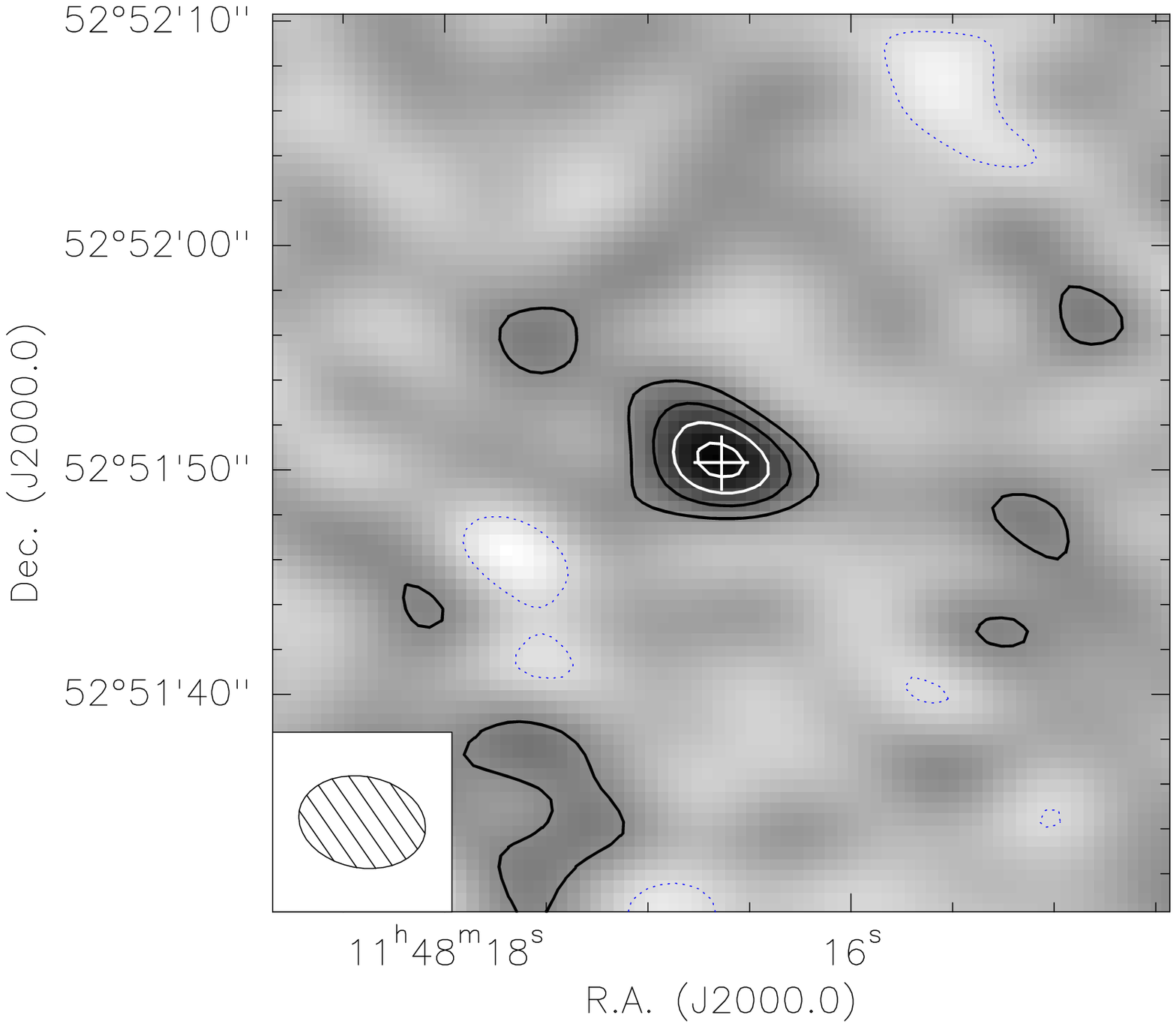}
\end{center}
\caption{Detections of CO emission in J1148 at $z=6.4$. {\bf Left panel:} VLA observations of the CO(2-1) transition. Reproduced from Fig. of 4 from Stefan et al. (2015). {\bf Middle panel:} CO(3-2) emission obtained with the same instrument. Reproduced from Fig. 1 of Walter et al. (2004) and reproduced by permission of the AAS. {\bf Right panel:} CO(6-5) emission detected with PdBI. Reproduced from Fig. 2 of Bertoldi et al. (2003b). Credit: Bertoldi Frank, A\&A, 409, 47, 2003, reproduced with permission \copyright ~ESO.
}
\label{fig:CO}
\end{figure*}
\begin{figure*}
\begin{center}
\includegraphics[height=2.4in]{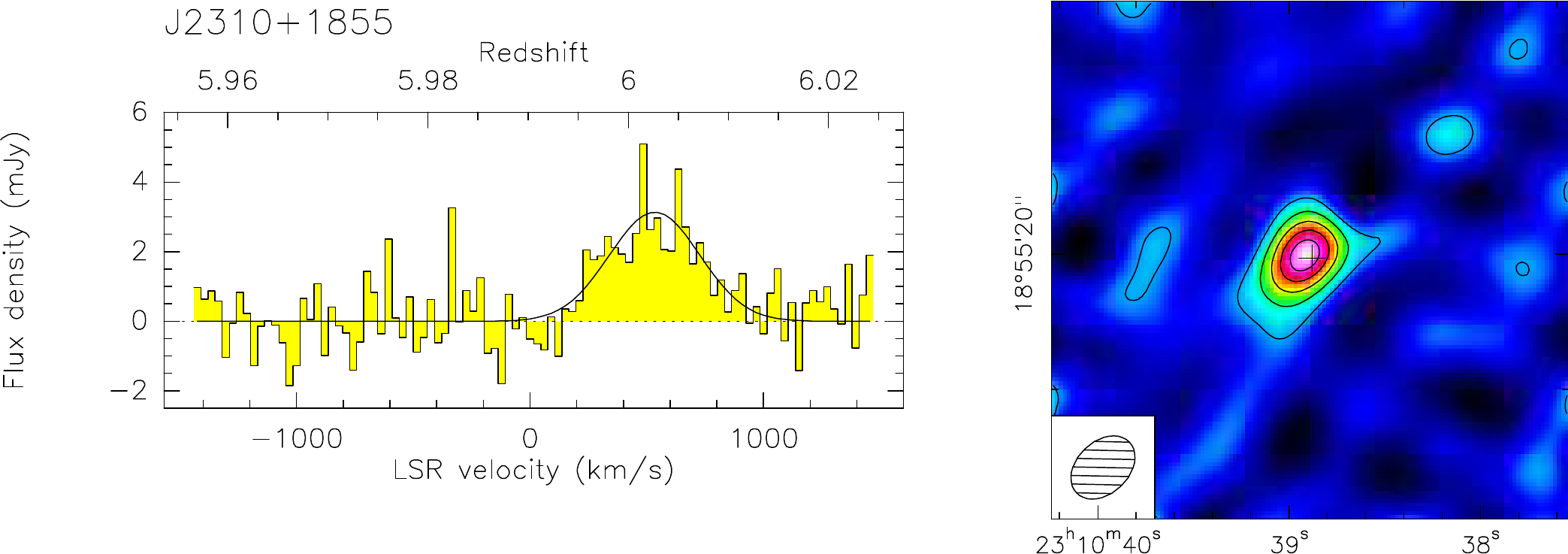}
\caption{PdBI observation of the CO (6-5) line from J2310+1855. The left panel shows the CO (6-5) line spectrum binned to 30 $\rm km\,s^{-1}$ channels. 
The solid line is a Gaussian fit to the line spectrum. The right panel shows the intensity map of the CO (6-5) line emission. The 1$\sigma$ rms noise of the map is $\rm 0.13\,Jy\,km\,s^{-1}\,Beam^{-1}$ and the contours in steps of 2$\sigma$. The beam size of $\rm 5.4''\times3.9''$ is plotted on the bottom left. The cross denotes the position of the optical quasar. Reproduced from Fig. 1 of Wang et al. (2013) and reproduced by permission of the AAS.}
\label{fig:CO_Wang}
\end{center}
\end{figure*}
Interestingly, these observations are consistent with the expectations of models and simulations of the formation of
massive galaxies in the early Universe accompanied by black hole growth. More specifically, models expect a similar very
high quasar-driven outflow rate (Fig.\ref{fig:feedb_models_valiante}, Valiante et al. 2012, while expecting a negligible contribution
from SN-driven winds) and similar outflow velocities and outflow extension (Fig.\ref{fig:feedb_models_costa}, Costa et al. 2015). 

Another interesting result, also going in the direction of reducing the efficiency of quasar
feedback, is that the outflow activity appears to be episodic/bursty and not continuous.
This could be observationally inferred
by measuring the dynamical time (distance from the centre divided by outflowing velocity)
of the individual outflowing clouds in J1148 (Cicone et al. 2015). This analysis enables the outflow activity
to be traced ``back in time''. The results are shown in Fig. \ref{fig:tdyn} for the evolution of the mass outflow rate, kinetic
power and momentum rate. Although various selection and sensitivity issues potentially affect the uncertainties associated
with these quantities,
the diagrams in Fig. \ref{fig:tdyn} suggest that the outflow has been active only for the last 30~Myr, and in a bursty mode.
\begin{figure*}
   \begin{center} 
   \includegraphics[width=6.4cm]{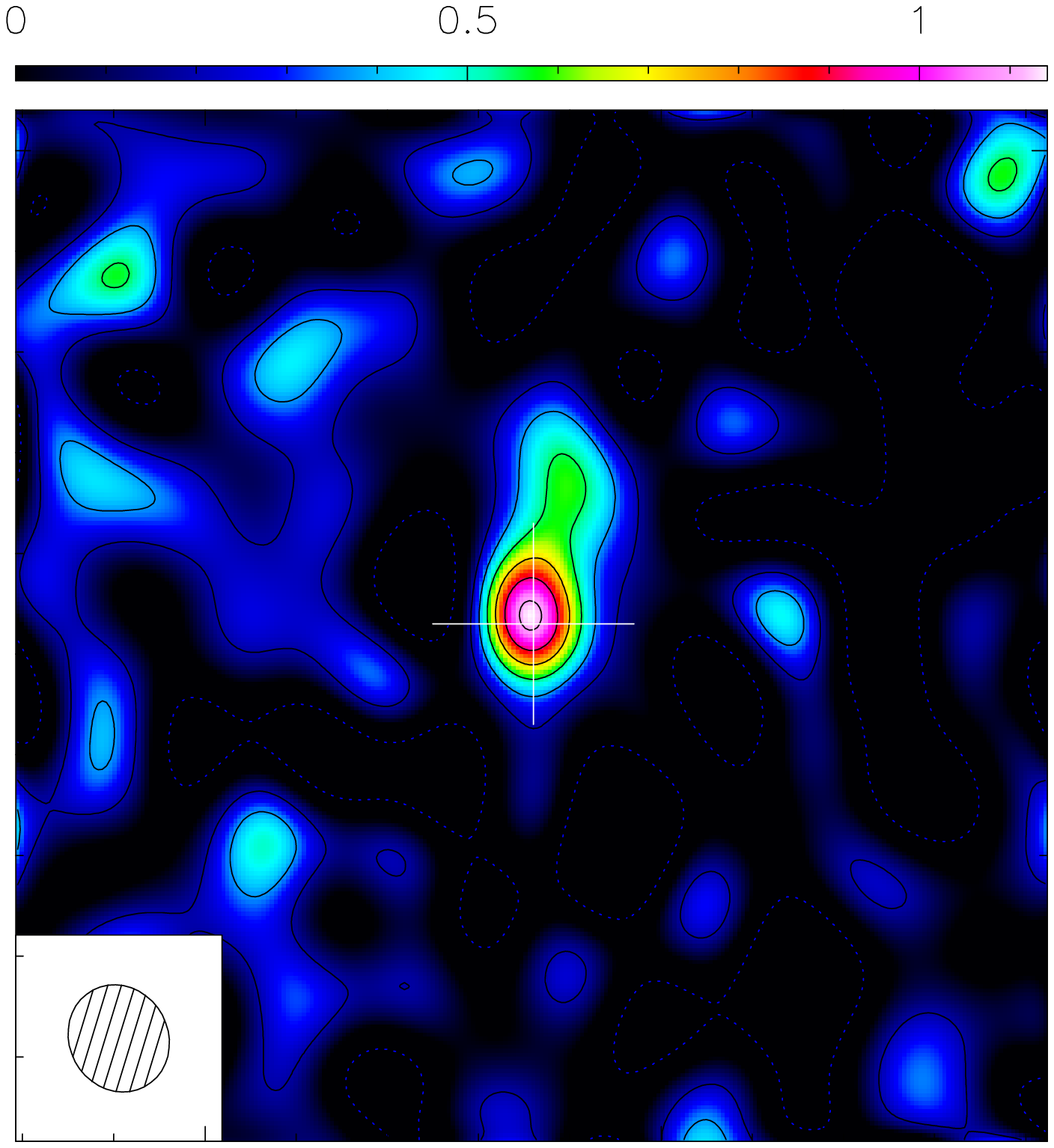}
\includegraphics[width=9.4cm]{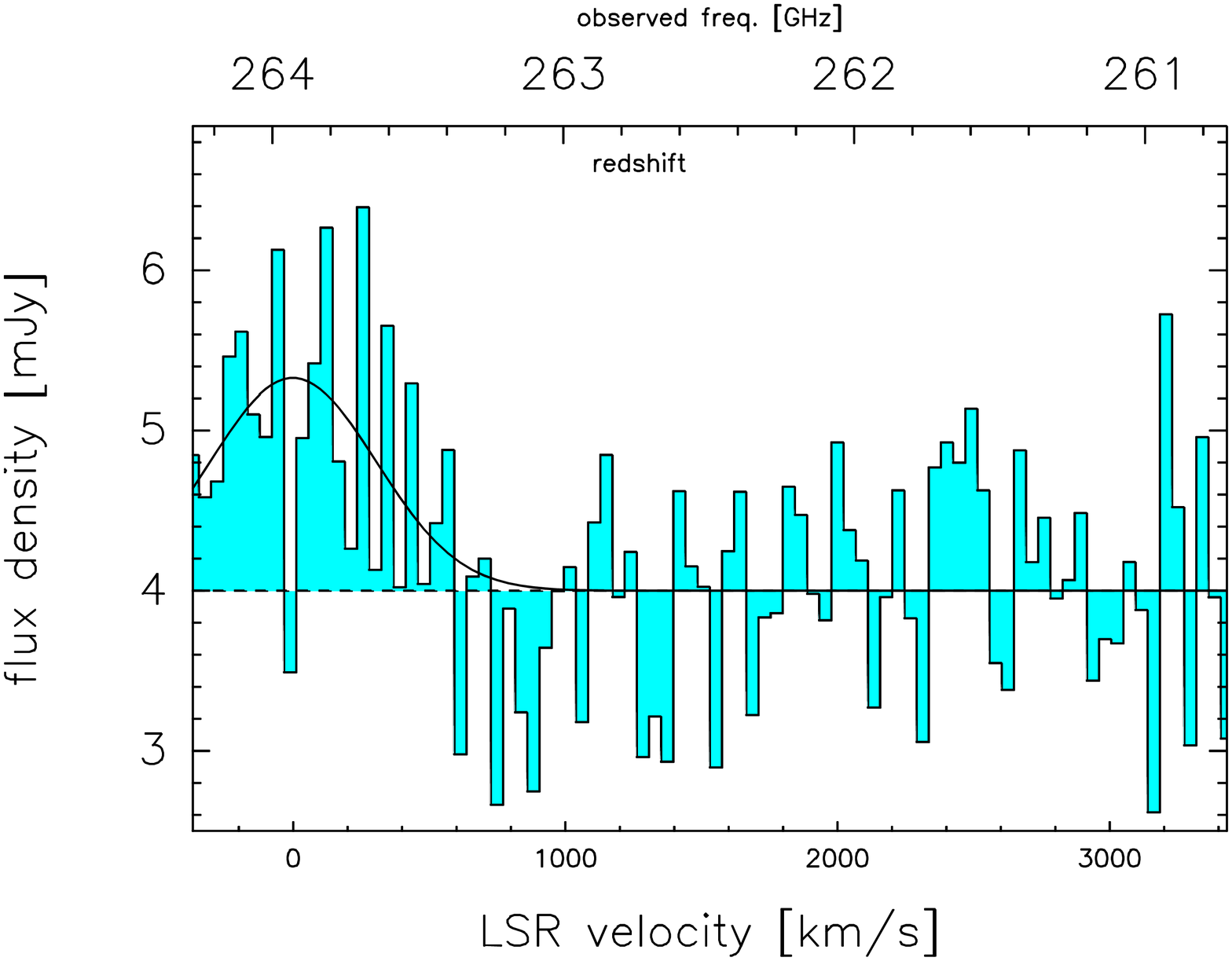}
   \end{center}
\caption{
{\bf Left panel:} PdBI CO(17--16) emission detected in J1148. The 1$\sigma$ noise in the images is 0.183\,mJy\,beam$^{-1}$ and contours are plotted from -1$\sigma$ to 6$\sigma$. The cross indicates the optical position of J1148. The beam ($1.1''\!\times\!0.98''$) is plotted in the lower left of the panel). {\bf Right panel:} Spectrum of the CO(17--16) line of J1148 as observed with the PdBI (channel width: 39\,MHz\,=\,44\,km\,s$^{-1}$, noise per channel: 0.8\,mJy), shown on the top of a 4.0$\pm$0.1~mJy continuum emission at 262 GHz. Adapted from Gallerani et al. (2014).}
\label{fig:tot_CO1716}
\end{figure*}
\section{Molecular gas in $z\sim 6$ quasar hosts}
The carbon monoxide (CO) is the most abundant molecule after molecular hydrogen ($H_2$) and its rotational transitions are excited through collisions with $H_2$. Thus, rest frame FIR lines arising from rotational transitions in the CO molecule can be used to measure the $H_2$ abundance (in the case of low-$J$ transitions) and to constrain the excitation conditions of molecular gas (in the case of high-$J$ transitions).
\subsection{Low-$J$ ($J\leq 7$) CO lines}
CO lines have been observed in several $z\sim 6$ quasars at different total angular momentum $J$. The first CO detections in a $z\sim 6$ quasar has been obtained by Bertoldi et al. (2003b) and Walter et al. (2003) in J1148. These authors detected the CO(6-5) and CO(7-6) with the PdBI and the CO(3-2) with the VLA, respectively (see right-most and middle panels of Fig. \ref{fig:CO}). In the same object, an upper flux limit for the CO (1-0) line was obtained from observations with the Effelsberg 100-meter telescope. More recently, Stefan et al. (2015) have detected the CO(2-1) transition (see the left-most panel of Fig. \ref{fig:CO}). The angular resolution of the  CO(2-1) and CO(3-2) lines is high enough to resolve the CO emission and provide estimates on the spatial extent of the $H_2$ distribution in this high-$z$ quasar that results to be $\leq 4~\rm kpc$. Moreover the intensity of the lines is consistent with $H_2$ masses $\sim 2\times 10^{10}~\rm M_{\odot}$. This large amount of molecular gas is common to high-$z$ quasars, since similar results have been obtained in $\sim$20 quasars at $5.7<z<6.2$ (Wang et al. 2010, 2011a, 2011b, 2013, see Fig. \ref{fig:CO_Wang}).
\subsection{High-$J$ ($J> 7$) CO lines}
\begin{figure*}
   \begin{center} 
   \includegraphics[width=17cm]{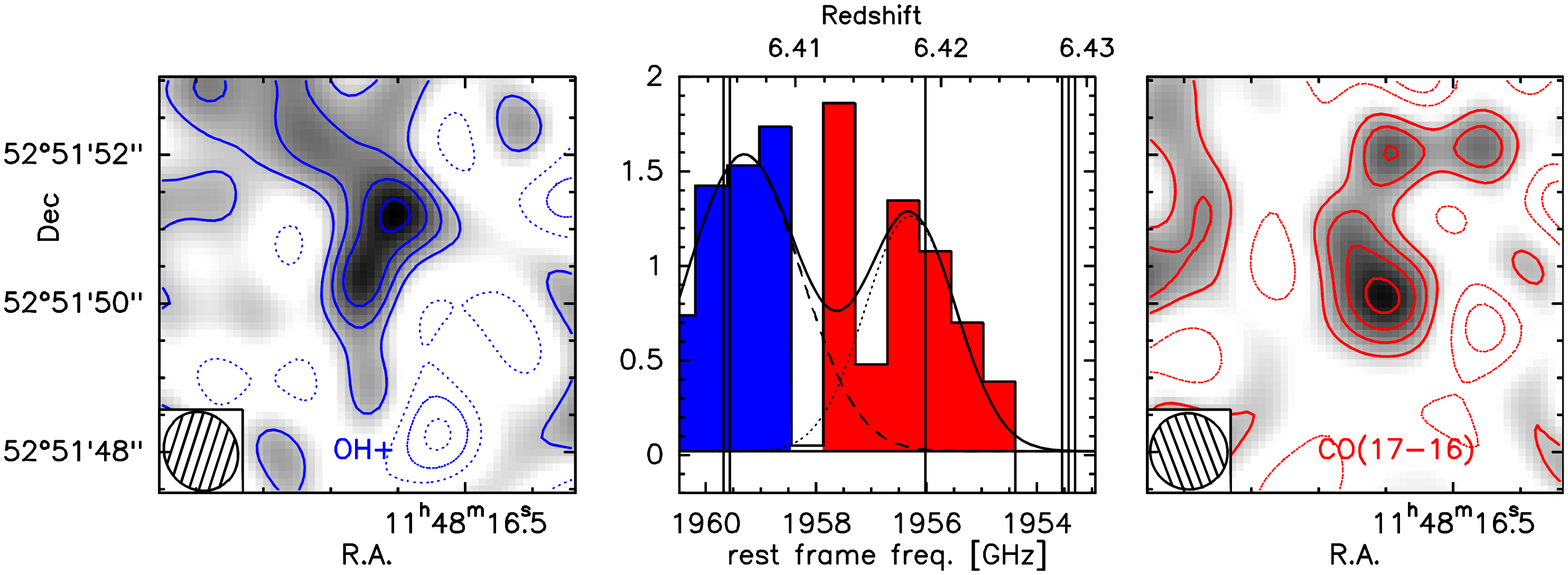}
   \end{center}
\caption{{\bf Middle panel}: Spectrum shown in Fig \ref{fig:tot_CO1716}, zoomed in the rest frame frequency range $1953<\nu_{RF}/[GHz]<1961$, assuming $z=6.4189$ for the J1148 redshift. Vertical lines show the frequencies of five OH$^+$ transitions and the CO(17-16) transition ($\nu_{\rm RF}=1956.018137$~GHz). {\bf Left panel}: OH$^+$ map obtained by integrating over the channels denoted by the blue shaded region in the middle panel. {\bf Right panel}: CO(17-16) map obtained by integrating over the channels denoted by the red shaded region in the middle panel.}
\label{fig:CO_OH}
\end{figure*}
High-$J$ CO lines arise from states $>100$~K above ground and have critical densities $>10^5~\rm cm^{-3}$. These lines trace the warm, denser molecular gas in the center of galaxies, and are difficult to excite solely with star formation. Thus, they can be used to test models that distinguish between AGN and starburst systems (e.g Meijerink et al. 2009; Schleicher et al. 2010a). More specifically, a quantity that is generally used to investigate the excitation conditions of the molecular gas is the so-called CO Spectral Line Energy Distribution (COSLED). This quantity is defined as the ratio of the ($J \rightarrow J-1$) CO rotational transition luminosity to a fixed CO transition. The COSLED represents a powerful tool to investigate the molecular gas kinetic temperature, $T_K$, that depends on the radiative heating due to stars and AGN (e.g. Obreschkow et al. 2009). More specifically, a higher $T_K$ boosts higher $J$ transitions, thus pushing the COSLED maximum towards larger $J$  values. The presence of high energy ($\geq 1~{\rm keV}$) photons emitted by an AGN causes the CO line intensities to rise well beyond $J=10$, making these highly excited lines powerful probes of quasar activity and sensitive tracers of X-ray Dominated Regions (XDR) (e.g. Schleicher et al. 2010a). 

High-$J$ CO lines are commonly detected in the local Universe up to $J=30$ (e.g. Panuzzo et al. 2010; Meijerink et al. 2013; Hailey-Dunsheath et al. 2012; Mashian et al. 2015). At higher redshifts, the highest-$J$ CO lines observed are the CO(17-16) in J1148 at $z=6.4$ (Gallerani et al. 2014) and the CO(11-10) in APM 08279+5255 at $z=3.9$ (Weiss et al. 2007).\\

While observing the dust continuum emission in J1148 with the PdBI receivers tuned at $\sim 262$~GHz, Gallerani et al. (2014) serendipitously detected strong line emission in the data, with a significance of 6.2$\sigma$ (see Fig. \ref{fig:tot_CO1716}). This strong unexpected emission has been ascribed to the sum of the CO(17--16) emission line ($\nu_{\rm RF}=1956.018137$~GHz), and five $\rm OH^+$ rotational transitions  ($1953.426326\leq \nu_{\rm RF}/\rm GHz\leq 1959.675435$) for a total velocity--integrated flux of the system of S$\Delta v=$1.01$\pm 0.16$ Jy\,km\,s$^{-1}$ (see Gallerani et al. 2014 for further details).\\ 
In the middle panel of Fig. \ref{fig:CO_OH}, the observed spectrum rebinned to 88 km~s$^{-1}$ is shown. The data have been fitted through a double Gaussian. The dotted line (shaded red region) represents the CO(17-16) line ($z_{\rm CO(17-16)}\sim 6.418$; FWHM$_{\rm CO(17-16)}\sim 297$\,km\,s$^{-1}$), while the dashed line (blue shaded region) denotes the OH$^+$ line ($z_{\rm OH^+}\sim 6.420$; FWHM$_{\rm OH^+}\sim 373$\,km\,s$^{-1}$). The relative contribution of the CO(17--16) line to the total emission results to be $\sim 40$\%,
 that corresponds to a CO(17--16) luminosity $L_{\rm CO(17-16)}=(4.9\pm 1.1)\times 10^8~L_{\odot}$ (Solomon et al. 1992).

\begin{figure}
\hspace{-0.7cm}
\includegraphics[width=9.2cm]{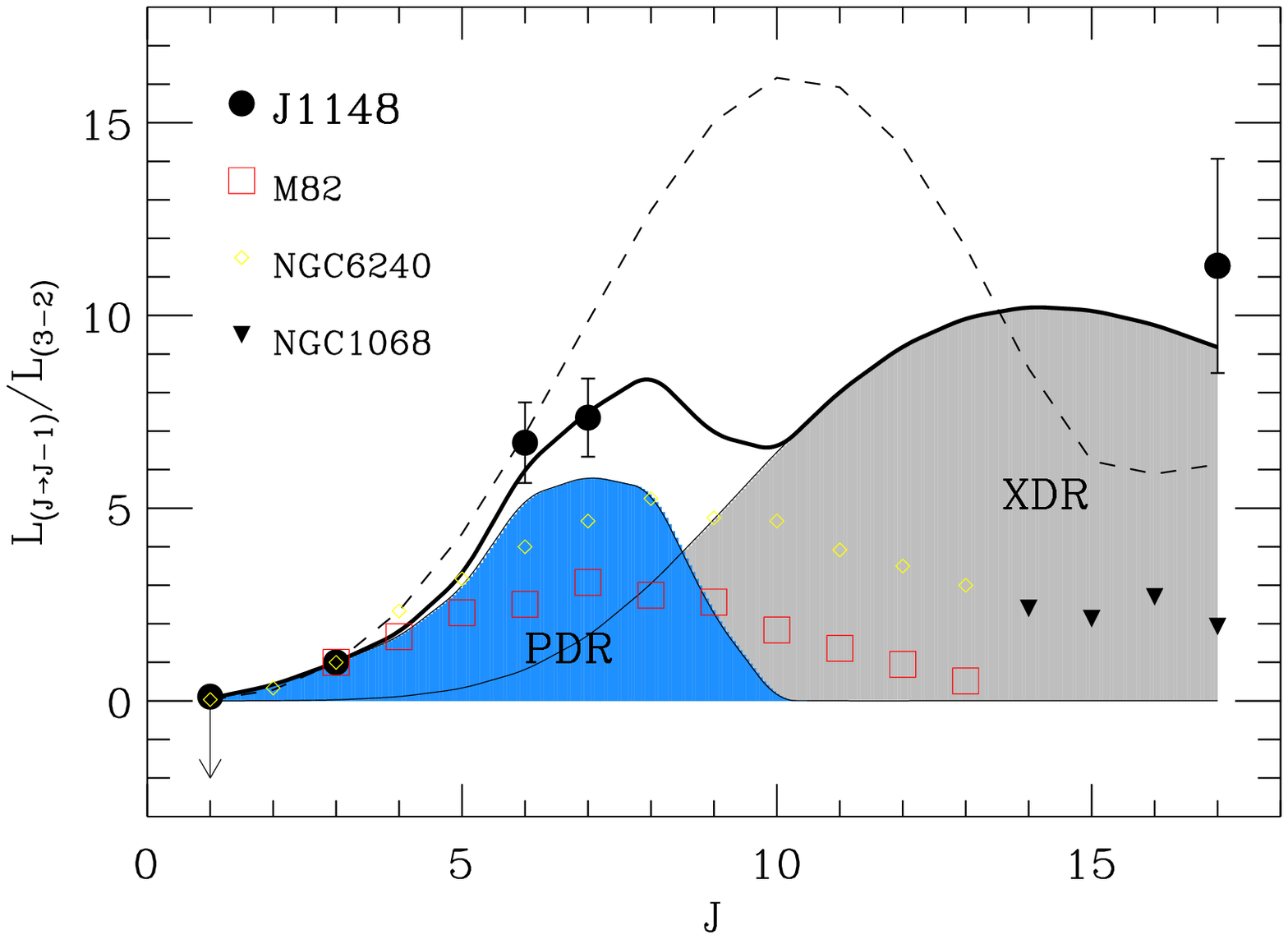}
\caption{CO Spectral Line Energy Distribution (COSLED) of J1148. Circles denote observations of several CO transitions: CO(1--0) (upper limit from Bertoldi et al. 2003b); CO(3--2) (Walter et al. 2003); CO(6--5) (Bertoldi 2003b); CO(7--6); CO(17--16) (Gallerani et al. 2014). The dashed line shows the best-fit obtained in the PDR model case, while the thick solid line represents the best-fitting composite model: the relative contributions from PDR and XDR are shown by the blue and gray shaded regions, respectively. The detection of the CO(17-16) line can not be reproduced by the PDR model alone, thus suggesting contribution from XDRs. Empty squares, empty diamonds, filled triangles show the COSLED observed in several local galaxies. Adapted from Gallerani et al. (2014).}
\label{fig:COSLED}
\end{figure}
By combining all the CO observations obtained in J1148 (see Table \ref{J1148}), it has been possible to construct its COSLED, shown in Fig. \ref{fig:COSLED} with filled circles, normalized to the CO(3--2) transition. These data have been interpreted by comparing them with results of radiative transfer calculations of Photo-Dissociation Regions (PDR, i.e. excited by UV radiation field) and X-ray Dominated Regions (XDR, i.e. excited by X-ray photons) developed by Meijerink et al. (2005; 2007). For this comparison, different gas densities ($n_{\rm PDR}$, $n_{\rm XDR}$ in ${\rm cm^{-3}}$), UV fluxes ($G_{\rm 0}$ in Habing units of ${\rm 1.6 \times 10^{-3} erg~s^{-1}cm^{-2} }$), and X-ray fluxes ($F_{\rm X}$ in ${\rm erg~s^{-1}cm^{-2}}$) at the cloud surface have been considered. 
\begin{table}
\caption{[CII] and CO line fluxes and width in SDSS\,J1148+5251. \label{J1148}}
\label{sample-table}
\begin{center}
\begin{tabular}{c c c c c}
\hline\hline
& $S_{\nu}{\rm d}v$& FWHM&Ref\\
&(Jy~km~s$^{-1}$)&(km~$s^{-1}$)&\\
\hline
\vspace {.2cm}
[CII] {\it narrow}&13$\pm$3&353$\pm$47&[1]$^a$\\
\vspace {.2cm}
[CII] {\it broad}&21$\pm$9&2119$\pm$706&[1]$^a$\\
\vspace {.2cm}
CO(1-0)& $<$0.11         &  --  &[2] \\ 
\vspace {.2cm}
CO(2-1)&0.09$\pm$0.01&298$\pm$26&[3]\\
\vspace {.2cm}
CO(3-2)& 0.20$\pm$0.02 & 320& [2][4]\\ 
\vspace {.2cm}
CO(6-5)& 0.67$\pm$0.08 & 279&[2] \\ 
\vspace {.2cm}
CO(7-6)& 0.63$\pm$0.06 & 297$\pm$35&[2][5]\\ 
\vspace {.2cm}
CO(17-16)&0.40$\pm$0.09&297&[6]\\
\hline\hline
\end{tabular}
\end{center}
\tabnote{References: [1] Cicone et al. (2015); [2] Bertoldi et al. (2003b)$^b$; [3] Stefan et al. (2016)$^c$; [4] Walter et al. (2003)$^b$; [5] Riechers et al. (2009); [6] Gallerani et al. (2014)\\
$^a$ We report the velocity-integrated flux for the narrow and broad component obtained when the [CII] spectrum is extracted with a circular aperture having a diameter 8" large (see Sec. 5 and Table 1 in Cicone et al. 2015).\\ 
$^b$ From Table 1 in Riechers et al. (2009)\\
$^c$ Stefan et al. (2016) reported a velocity-integrated flux of 94.6$\pm$7.7 mJy beam$^{-1}$~km~s$^{-1}$, for a beam size 1.''07 × 0.''97. 
}
\end{table}

The result of this comparison is that PDR models alone (dashed line in Fig. \ref{fig:COSLED}) can not fairly explain the observed high-J (J$\geq$7) CO line observations. 
Vice versa, the COSLED predicted by a model in which the molecular cloud (MC) emission is the sum of the contribution from a higher-density XDR region embedded in a more rarefied and extended PDR envelope (solid line in Fig. \ref{fig:COSLED}) is in very good agreement with observed data. According to this model, individual MCs have a typical mass $M_{\rm c}\sim 2\times 10^5~{\rm M_{\odot}}$ and a radius $r_{\rm c} \sim 10$ pc; the XDR core ($n_{\rm XDR}=10^{4.25\pm 0.25}$) is irradiated by an X-ray flux $F_{\rm X}=160\pm 70$, while the FUV flux at the PDR surface ($n_{\rm PDR}=10^{3.25\pm 0.25}$) is $G_{\rm 0}=10^{4.0\pm 0.25}$. 

\begin{figure}
\vspace{-1.2cm}
\hspace{-1.2cm}
\includegraphics[height=13.8cm]{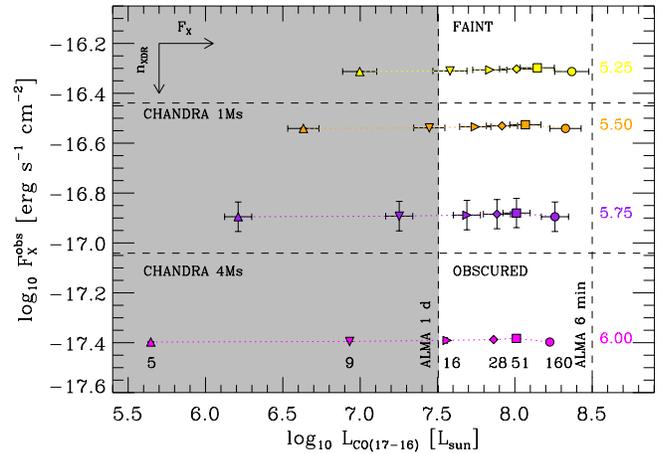}
\vspace{-6.5cm}
\caption{Observed soft X-ray flux ($F_{\rm X}^{\rm soft}$) versus CO(17--16) luminosity ($L_{\rm CO(17-16)}$) computed in the case of a $z=7$ quasar, powered by a $3\times 10^{6}~\rm M_{\odot}$ black hole, and considering an obscuring gas column density $N_{\rm H}>N_{\rm H}^*=10^{24}$ cm$^{-2}$. Symbols and colors show different interstellar medium properties of the host galaxy: yellow, orange, violet, and magenta symbols refer to increasing XDR densities in the range $10^{5.25}<n_{\rm XDR}/\rm cm^{-3}<10^{6}$; triangles, downwards triangles, rightwards triangles, diamonds, squares, circles represent increasing X-ray fluxes at the illuminated surface of the XDR in the range $5<F_X<160 ~\rm erg~s^{-1}cm^{-2}$. Error bars have been computed by considering the contribution of different PDR densities ($10^3<n_{\rm PDR}/\rm cm^{-3}<10^{3.5}$) to the $L_{\rm CO(17-16)}$ luminosity. Adapted from Gallerani et al. (2014).}
\label{fig:ALMA_CHANDRA}
\end{figure}

The most straightforward interpretation of this result is that the detection of a prominent $J\geq 7$ line in high redshift objects strongly supports the presence of AGN activity. According to this extremely appealing possibility, high-J CO lines could allow to identify the progenitors ($M_{\bullet}\sim 10^{6-7} M_{\odot}$) of SMBHs observed in $z\sim 6$ quasars. In Fig. \ref{fig:ALMA_CHANDRA}, the expected X-ray flux of the elusive SMBH progenitors is compared with their expected CO(17-16) line luminosity. Given that the luminosity of the CO(17-16) line in these objects can be as high as $10^{7.5}\leq L_{\rm CO(17−16)}/L_{\odot} \leq 10^{8.5}$, they should be detectable with the ALMA full array in an observing time of 12 hr$\geq$ t $\geq$ 10 min.
\section{Summary and future prospects}\label{summary_sec}
In this paper, we have reviewed the physical properties of the highest redshift ($z\sim 6-7$) quasars known so far. First of all, we have discussed the formation and evolution scenarios of the black holes that power high-$z$ quasars, with special emphasis on the formation of massive ($\sim 10^3-10^5 \, \mathrm{M_{\odot}}$) black hole seeds through the direct collapse black hole (DCBH) mechanism. Then, we have focused on several properties of high-$z$ quasars as inferred by submm and mm observations. Here below, we summarize the main results.
\begin{enumerate}[label=(\roman*)]
\item {\bf Formation and evolution of SMBH seeds}\\ 

The DCBH channel is nowadays object of great interest in the community, 
above all because DCBH observational properties may be tested with current or upcoming instruments. In this context, we have discussed the case of CR7, a $z=6.6$ Lyman-alpha emitter that has been indicated as a very good DCBH candidate by several studies in the last year. In particular, its Ly$\alpha$ and HeII luminosity are compatible with a BH born from a DCBH seed (Pallottini et al. 2015). In the future, deeper X-ray observations and/or variability studies of this source may reveal its real nature. Moreover, very recently, a photometric method to select DCBH candidates in deep multi-wavelength fields has been proposed. This method has already identified a couple of DCBH candidates, at redshift $z > 6$;
its extension to additional multi-wavelength fields will surely lead to the identification of more DCBH candidates. Nonetheless, only a spectroscopic study of the pre-selected sources will be able to confirm their real nature.\\

\item {\bf Dust continuum emission}\\

Dust continuum emission studies have revealed that high-$z$ quasars have FIR luminosities $L_{\rm FIR}\sim 10^{12}-10^{13}\rm L_{\odot}$ that correspond to dust masses $\sim 10^8\rm M_{\odot}$ and star formation rates of few hundred up to few thousands $\rm M_{\odot}~yr^{-1}$. These studies have also shown that the FIR activity of high-$z$ quasars shows a mild correlation with their bolometric luminosity ($L_{\rm FIR}\propto L_{\rm bol}^{0.6}$), and is higher than the FIR activity of local quasars, suggesting that the growth of massive early black holes is accompanied by an higher level of star formation activity with respect to low-$z$ systems.\\
\item {\bf [CII]158$\mu$m emission lines}\\

High-$z$ quasar host galaxies are characterized by [CII] luminosities that range from $10^8 \rm L_{\odot}$  to $10^9 \rm L_{\odot}$, and [CII] lines are characterized by $FWHM\sim 200-500~\rm km~s^{-1}$. In some cases, it has been also possible to resolve the [CII] emission and thus to measure the size of the region from which it is arising and its dynamical mass ($M_{\rm dyn}\sim 10^{10}-10^{11}\rm M_{\odot}$). These measurements permit to study the co-evolution of SMBHs and galaxies at early epochs finding that black hole growth precedes galaxy assembly. Thus, quasars with the largest black holes are not necessarily in the most massive galaxies in early epochs. Results discussed here are based on the earliest ALMA data in which the [CII] emission is resolved, often only marginally, in few cases. Increased sensitivity and spatial resolution of future ALMA data will allow us to fully resolve the [CII] emission in a large sample of high-redshift quasars and to obtain detailed dynamical modeling of individual objects. Improved constraints on both [CII] emission sizes and velocity line widths, will make estimates of the dynamical mass of the host galaxy more accurate and will finally help us understanding how the $M_{BH}-M_{dyn}$ relation does evolve with redshift.\\

\item {\bf Quasar-driven outflows}\\

The co-evolution of SMBHs and host galaxies is possibly shaped by quasar feedback. In particular, quasar-driven outflows are considered as possible mechanisms to quench star formation in massive galaxies that consequently evolve into local massive elliptical galaxies. The high [CII] line luminosity ($>10^{10}\rm L_{\odot}$) in J1148 ($z=6.4$) has allowed Maiolino et al. (2012) and Cicone et al. (2015) to obtain the first evidence of quasar-driven outflow in this system. The inferred outflow rate ($1400\pm 300~\rm M_{\odot}~yr^{-1}$) is of the same order of the star formation rate, thus implying a short gas depletion time ($<10~\rm Myr$). This result suggests that the star formation in the host galaxy of J1148 would be quenched on very short timescales.\\ 
Interestingly, the kinetic power of this $z\sim 6$ quasar outflow is only a tiny fraction of the AGN luminosity ($\sim$0.05\%), implying that this specific quasar is not particularly efficient in driving the outflow, especially compared with local quasars (Cicone et al. 2014) and compared with model expectations (King 2010), according to which quasar-driven outflows can reach a kinetic power as high as 7\% of the AGN radiated luminosity. Therefore, it is likely that other quasars at $z\sim 6$ are characterized by much more powerful outflows, which can possibly be much more efficient in quenching star formation. More ALMA observations are being obtained on more quasars at high redshift with the aim of looking for broad wings in the shape of the [CII] emission line. These new data will be used to unveil the origin and nature of quasar-driven, high-$z$ outflows. As in the case of local quasar studies, the inferred outflow rate and kinetic power of these objects will be related to the SFR of their host galaxies and to the AGN luminosity. For a fair comparison with the local Universe, observations of outflows in a tens of $z\sim 6$ quasars are required.\\

\item {\bf Carbon monoxide (CO) emission lines}\\

We have recalled the molecular hydrogen properties of $z\sim 6$ quasars as inferred from CO emission line studies. VLA and PdBI observations of low-J ($J\leq 7$) CO lines have shown that these systems are characterized by large $H_2$ masses ($\sim 10^{10}~M_{\odot}$), concentrated on relatively small scales ($\leq 4 ~\rm kpc$). Also in this context, the most surprising result has been obtained in J1148; in fact, Gallerani et al. (2014) have observed in this object the most excited CO line ever detected at high-$z$, namely the CO(17-16) emission line. The excitation of high-$J$ CO lines have been ascribed by these authors and also by previous works (e.g. Meijerink et al. 2007; Schleicher et al. 2010a) to X-ray Dominated regions. This outlines the exciting possibility of detecting SMBH ancestors through observations of high-$J$ CO lines.\\
Nevertheless, the interpretation of these lines might be more complex. First of all, the limited number of CO lines detected so far hampers a full characterization of the uncertainties/degeneracies of the PDR and XDR model parameters. Moreover, Herschel observations have shown that other processes (e.g. shocks; Panuzzo et al. 2010; Meijerink et al. 2013; Hailey-Dunsheath et al. 2012) might play a role in the excitation of high-J CO lines. Thus, to further test the possibility of using high-$J$ CO lines as tools for detecting AGN activity, it will be crucial to understand the relative contribution of X-rays and shocks to the COSLED.\\ 
With the end of the Herschel mission, it will be very challenging to improve our interpretation of high-J lines from observations in the local Universe. Vice versa, ALMA could in principle easily detect these lines at high-$z$. Additional observations of high-J CO lines ($15<J<19$) obtained in the ALMA Cycle 2 in other $z\sim 6$ quasars will add further constraints on the physical processes responsible for the molecular gas excitation. 
\end{enumerate}
\subsection{A forward look to future observations}
The last two decades of optical/mm observations have revealed that $z\sim 6$ quasars are powered by $\sim 10^8-10^{10}~\rm M_{\odot}$ black holes, hosted by and co-evolving with highly star forming ($SFR\sim 100-1000~\rm M_{\odot}yr^{-1}$), massive ($M_{dyn}\sim 10^{10}-10^{11}~\rm M_{\odot}$) galaxies, rich of gas  ($M_{H2}\sim 10^{10}~\rm M_{\odot}$), dust  ($M_{dust}\sim 10^8~\rm M_{\odot}$), and metals ($Z\gtrsim 0.3~\rm Z_{\odot}$).

On the one hand, merging galaxies and/or gas accretion are expected to provide the main mechanisms for BH seeds to grow up into SMBHs; on the other hand, quasar-driven feedback may prevent the fuelling of gas into the central SMBH and/or enhance the gas density in those regions crossed by the outflow, thus quenching and/or triggering star formation in the host galaxy. Both the formation and evolution of SMBHs and their impact on the host galaxies certainly represent the most important open issues in our understanding of the physics of $z\sim 6$ quasars.

Further information on the nature of these rare objects and their progenitors will be obtained through deep and high resolution ALMA observations (as discussed in (i)-(v); see also Schleicher et al. 2010b), but also by means of future infrared (e.g. JWST\footnote{James Webb Space Telescope (e.g. Agarwal et al. 2013, Natarajan et al. 2016)}, SPICA\footnote{SPace Infrared telescope for Cosmology and Astrophysics (Spinoglio et al., in preparation)}) and X-ray (e.g. Athena\footnote{Advance Telescope for High Energy Astrophysics (e.g. Barret et al. 2013)}, Lynx\footnote{Previously known as X-ray Surveyor, NASA mission}) space missions, and optical/infrared (e.g. E-ELT\footnote{European Extremely Large Telescope (e.g. Maiolino et al. 2013)}), radio (e.g. SKA\footnote{Square Kilometer Array (e.g Mellema et al. 2013, Manti et al. 2016)}) ground based telescopes. 

\section*{Acknowledgement}
RM acknowledges support from the ERC Advanced Grant 695671 ``QUENCH'' and from the Science and Technology Facilities Council (STFC). FP acknowledges the NASA-ADAP grant nr. MA160009. We thank the anonymous reviewer for her/his careful reading of our manuscript and her/his insightful comments and suggestions. We are grateful to Eduardo Ba$\rm \tilde{n}$ados, Claudia Cicone, Giorgio Lanzuisi, Jianwei Lyu, Benny Trakhtenbrot, Rosa Valiante, Bram Venemans for useful comments on the manuscript.

\nocite*{}
\bibliographystyle{pasa-mnras}
\end{document}